\documentclass[twocolumn]{aastex62}
\usepackage{mathtools}
\usepackage{amsmath}
\usepackage{subfigure}
\usepackage{autobreak}
\usepackage{breakurl}

\shorttitle{Galaxy cluster catalog}
\shortauthors{J. Gao et al.}

\begin{document}

\title{A CATALOG OF GALAXY CLUSTERS IDENTIFIED FROM SCUSS, SDSS, AND UNWISE}

\correspondingauthor{Hu Zou}
\email{zouhu@nao.cas.cn; jinghuabnu@nao.cas.cn}

\author{Jinghua Gao}
\affil{Key Laboratory of Optical Astronomy, National Astronomical Observatories, Chinese Academy of Sciences, Beijing 100012, China}
\author{Hu Zou}
\affil{Key Laboratory of Optical Astronomy, National Astronomical Observatories, Chinese Academy of Sciences, Beijing 100012, China}
\author{Xu Zhou}
\affil{Key Laboratory of Optical Astronomy, National Astronomical Observatories, Chinese Academy of Sciences, Beijing 100012, China}
\author{Xu Kong}
\affil{Key Laboratory for Research in Galaxies and Cosmology, Department of Astronomy, University of Science and Technology of China, Hefei 230026, China}

\begin{abstract}
This paper presents the identification of galaxy clusters from the photometric redshift catalog based on three imaging surveys of SCUSS, SDSS, and unWISE. By applying a fast clustering algorithm, we obtain a total of 19,610 clusters in the redshift range of $0.05 < z < 0.65$ over a sky area of about 3,700 deg$^2$ in the south Galactic gap. Monte Carlo simulations show that the false detection rate is about 8.9\%. The redshift uncertainty is estimated to be about 0.013. The mass and richness of detected clusters are derived through the calibration based on the measurements of X-ray emission and Sunyaev--Zel'dovich effect. The median mass is $1.2\times10^{14} M_\sun$. 

\end{abstract}


\section{Introduction} \label{sec:intro}
As the largest gravitationally bound systems, galaxy clusters encode rich information of the universe. Most matter ($\sim80$\%) of galaxy clusters is in form of dark matter, the luminous matter ($\sim3-5\%$) is in galaxies, and the rest ($\sim15-17\%$) is in diffuse hot gas \citep{fer2012}. Luminous matter of galaxy clusters can trace the large-scale structure of the universe so that galaxy clusters provide ideal laboratories to study the relation between galaxies and their environments. Large samples of galaxy clusters have supported a lot of studies on cosmology \citep[e.g.,][]{gla2007,roz2010} and galaxy formation and evolution \citep[e.g.,][]{hao2011,gu2016,cer2019,yoo2019}. 

The overdensity feature of galaxy clusters is so evident that it can be easily detected in a single optical-band image \citep{abe1958,abe1989}. However, it is difficult to separate foreground and background galaxies in a single band. The member galaxies of such detected clusters might be contaminated due to the projection effect. In the past two decades, a number of wide or deep sky surveys have supported the detections of large samples of galaxy clusters \citep{yok2000, ham2001, des2005, sku2006, sco2007, wri2010, dej2013, aih2018}. The multi-band observations provide helpful information to separate galaxies in line of sight, so the clustering feature of galaxies can be easily detected. Galaxy clusters can be also detected in the X-ray and radio observations. The intracluster medium is heated by adiabatic compression and shocks to emit X-ray photons via thermal bremsstrahlung and line emission \citep{voi2005}. The cosmic microwave background (CMB) photons passing through the hot intracluster medium can be excited by inverse-Compton scattering and produce a characteristic spectral distortion to the CMB, known as the Sunyaev--Zel'dovich (SZ) effect \citep{sun1972, ble2015}. The detection of galaxy clusters based on the observations of X-ray emission and Sunyaev--Zel'dovich effect are independent on redshift measurements. The mass of galaxy clusters can be reliably determined by the weak-lensing measurements. Only a few massive clusters have such mass estimations \citep{wen2015}. However, there are a plenty of galaxy clusters with reasonable mass estimations from the X-ray emission and SZ effect \citep[e.g.,][]{ kra2006, pra2009,hoe2012, zha2013, has2013}, although the mass proxies may be biased due to uncertain assumptions such as hydrostatic equilibrium and spherical symmetry. These clusters can be used to calibrate the properties of galaxy clusters identified in optical observations \citep{sza2011, wen2015, wen2018}.

The detection methods based on multi-band imaging data have been the most efficient ways to obtain large samples of galaxy clusters. There are a number of studies about the identifications of galaxy clusters with multi-band optical and infrared imaging data. The most popular cluster catalogs are based on the Sloan Digital Sky Survey \citep[SDSS][]{yok2000}, such as maxBCG \citep{koe2007}, GMBCG \citep{hao2010},  AMF\citep{sza2011,ban2018}, WHL2012 \citep{wen2012}, and redMaPPer \citep{ryk2014}. \citet{koe2007} identified 13,823 clusters at $0.1 < z < 0.3$ using the maxBCG red-sequence method with SDSS DR5 data. \citet{hao2010} applied a GMBCG algorithm to SDSS DR7 data and selected 55,424 rich clusters at $0.1<z<0.55$.  \citet{sza2011} extracted a catalog of 69,173 clusters at $0.045<z<0.78$ from SDSS DR6 using an adaptive matched filter cluster finder and the photometric redshifts provided by \citet{oya2008}. \citet{wen2012} identified the largest catalog of 132,484 clusters in the redshift range of $0.05<z<0.8$ using the photometric redshifts of galaxies from SDSS III. \citet{ryk2014} applied the red-sequence cluster finder, redMaPPer, to SDSS DR8 data and presented a catalog of $\sim$25,000 clusters over the redshift range of $0.08 < z < 0.55$. These cluster-finding algorithms are based on either red-sequence feature of clusters or the overdensity feature related to photometric redshifts. The red-sequence methods rely on the color-magnitude relation of E/S0 galaxies and the existence of a brightest cluster galaxy (BCG) located close to the cluster center. The overdensity methods depend on the photometric redshifts and can detect those galaxy clusters not presenting obvious red-sequence features.  

Combining the data from South Galactic Cap u-band Sky Survey \citep[SCUSS;][]{zhou2016,zou2016}, SDSS, and unWISE \citep{lan2014,lan2016}, we have constructed an accurate photometric redshift (photo-z) catalog. Our catalog includes about 23.1 million galaxies with $r < 22$ mag, covering the redshift range of $z < 0.8$. The photo-z accuracy is about 0.02. The systematic bias between photometric and spectroscopic redshifts is reduced much compared with those photometric redshift catalogs that are only based on SDSS $ugriz$ data. In this paper, we plan to apply a new fast overdensity finding algorithm to the photo-z catalog and supply a new catalog of galaxy clusters. 
 
The structure of the paper is arranged as follows. Section \ref{sec:data} briefly describes the photometric data and corresponding photometric redshift catalog. The K-corrections and  absolute magnitudes for galaxies are derived in this section. Section \ref{sec:method} presents the cluster-finding algorithm and its performance in detail. The mass and richness of the detected clusters are estimated in this section. Section \ref{sec:catalog} describes the cluster catalog as well as comparisons with other catalogs. Section \ref{sec:sum} gives the summary. Throughout this paper, we assume a $\mathrm{\Lambda CDM}$ cosmology with $\Omega_m =0.3$, $\Omega_{\Lambda} =0.7$, and $H_0=70$ km  s$^{-1}$  Mpc$^{-1}$.

\section{Photometric redshifts and K-correction} \label{sec:data}
\subsection{The photometric redshift catalog}
Based on multi-wavelength photometric data from SCUSS, SDSS, and unWISE, \citet{gao2018} constructed a photometric redshift catalog covering about 3,700 deg$^2$ in the south Galactic cap. A total of 7 bands were used in the photometric redshift estimation, including SCUSS $u$, SDSS $griz$, and WISE $W1W2$. 
The forced model photometry is used, which provides unbiased colors of galaxies. All model magnitudes are measured according to the SDSS $r$-band model shape parameters and PSF profiles in different bands. The SDSS $u$ band is replaced with the SCUSS $u$, which is about 1.2 mag deeper.  The unWISE creates new coadds of the official WISE images and makes forced photometry on the coadds with consistent model parameters of galaxies from SDSS \citep{lan2014,lan2016}. The unWISE photometry is more complete than the official one. SCUSS $u$ band and WISE $W1$ and $W2$ bands help to reduce the systemic error relative to those photometric redshift estimations that are only based on the SDSS imaging data. We have estimated the photo-zs for about 23.1 million galaxies with $r$ band down to 22 mag and redshift up to 0.8. The average bias of $\Delta z_\mathrm{norm} = \frac{z_\mathrm{phot} - z_\mathrm{spec}}{1+z_\mathrm{spec}}$ is estimated to be 2.28$\times10^{-4}$, where $z_\mathrm{phot}$ and $z_\mathrm{spec}$ are photometric and spectroscopic redshifts, respectively. The photo-z accuracy, defined as the standard deviation of $\Delta z_\mathrm{norm}$, is about 0.019. 

As shown in Figure 5 and 7 of \citet{gao2018}, the photo-z bias becomes larger at $z_\mathrm{spec} > 0.6$ and $r > 21$ and the photo-z accuracy is better than 0.04 at $r < 21$ and goes up to 0.08 at $r = 22$. The number of photometric bands (``n\_filter") used for estimating the photo-z is also another key factor that determines the photometric redshift quality. For n\_filter $=$ 5, 6, and 7, the photo-z accuracy is 0.061, 0.028, and 0.019, respectively. We set some limits to the photometric redshift catalog to ensure the photo-z quality as shown below:
	 \begin{equation}
	 \begin{split}
	  r < 21.5, \\
	 0.05 < z_\mathrm{phot} < 0.65,  \\
	 \mathrm{n\_filter}>5.
	 \end{split}
	 \label{equ-limits}
	 \end{equation}
We restrict our cluster identification with a low redshift cut of $z_\mathrm{phot}> 0.05$, because the clusters at $z < 0.05$ have been effectively detected by both photometric and spectroscopic data. After the constraints of Equation (\ref{equ-limits}) are applied, about 40\% of galaxies in the photo-z catalog are eliminated. A total of about 14 million galaxies  are used for detecting clusters in this paper. 

\subsection{K-correction}
The K-corrections and absolute magnitudes are derived by using the method and code (kcorrect in IDL\footnote{\url{http://kcorrect.org/}}) of \citet{bla2007}. The version of kcorrect is v4\_3. This code uses a technique of nonnegative matrix factorization to produce nonnegative templates from the Bruzual-Charlot stellar evolution synthesis models of galaxies. The templates are then used to fit the observed spectral energy distributions of galaxies and calculate the K-corrections. Following \citet{wen2015}, the $r$-band luminosity of a galaxy is calculated in units of  $L^{*}(z)$, which is the evolved characteristic luminosity defined as $L^{*}(z)=L^{*}(0)*10^{0.4Qz}$ with $Q=1.16$ and $M^*_r(0.1)= -20.44$ taken from \citet{bla2003}.

\section{Cluster detection} \label{sec:method}

\subsection{Algorithm description}
There are many methods to find the clustering features of galaxies, however, the difficulty is to distinguish the cluster members from the foreground and background galaxies. In this paper, we use accurate photometric redshifts to relieve the projection effect and use a new fast clustering algorithm, Clustering  by Fast Search  and  Find  of  Density  Peaks (CFSFDP), to detect the overdensities. The CFSFDP algorithm was first proposed by \citet{rod2014}. It is based on the idea that the cluster center is characterized by both a higher density than their neighbours and a relatively large distance from others with higher densities. The algorithm can effectively detect clusters regardless of their shapes and automatically exclude outliers (e.g. foreground and background galaxies).  The specific procedure of our cluster detection using photo-z and CFSFDP algorithm is shown as below. 

\begin{enumerate} 

\item The local density ($\rho$) of each galaxy at a given redshift $z$ is calculated, which is defined as the number of galaxies within a radius of 0.5 Mpc and a photometric redshift bin of $z \pm 0.04(1+z)$. The redshift bin is set twice larger than the photo-z accuracy so that 95.4\% of  cluster members can be selected. 

\item The shortest distance ($\theta$ in Mpc) from a given galaxy to any other galaxies with higher local densities in the redshift bin are measured. The $\theta$ values of galaxy members in a cluster should be small enough, except for the density peak of the cluster. 

\item The center candidate of a cluster is defined as the position of the galaxies with the highest density, where plenty of member galaxies are concentrated within a radius of 0.5 Mpc. The local density $\rho$ at the peak should be significantly higher than the field galaxies. The average density $\rho_\mathrm{field}$ of the field galaxies along the redshift is measured over the entire footprint. At a specified redshift, the field density is computed within the redshift bin of $z \pm 0.04(1+z)$ and scaled to the circular area with a radius of 0.5 Mpc. We require that the potential cluster center should have $\rho > 5\rho_\mathrm{field}$, which is high enough to present the overdensity feature. The left panel of Figure \ref{fig-densityline} shows the galaxy distribution in the plane of $\rho$ vs. $z$.  The solid line presents 5 times $\rho_\mathrm{field}$. 

\item The cluster center should also have a large enough value of $\theta$ so that it is located at a local density peak and meanwhile far away from other peaks. According to the galaxy cluster catalog of \citet{wen2012}, 99\% of their clusters have $r_{200}$ larger than 0.75 Mpc, where $r_{200}$ is the radius within which the mean mass density is 200 times that of the critical cosmic mass density. We set a looser limit of 1.5 Mpc to $\theta$. Thus, the candidates of cluster centers are those galaxies with $\theta > 1.5$ Mpc and $\rho > 5\rho_\mathrm{field}$. The right panel of Figure \ref{fig-densityline} shows the galaxy distribution on the plane of $\rho$ vs. $\theta$. The blue points represent for the cluster center candidates. 

\begin{figure*}[!ht]
\plottwo{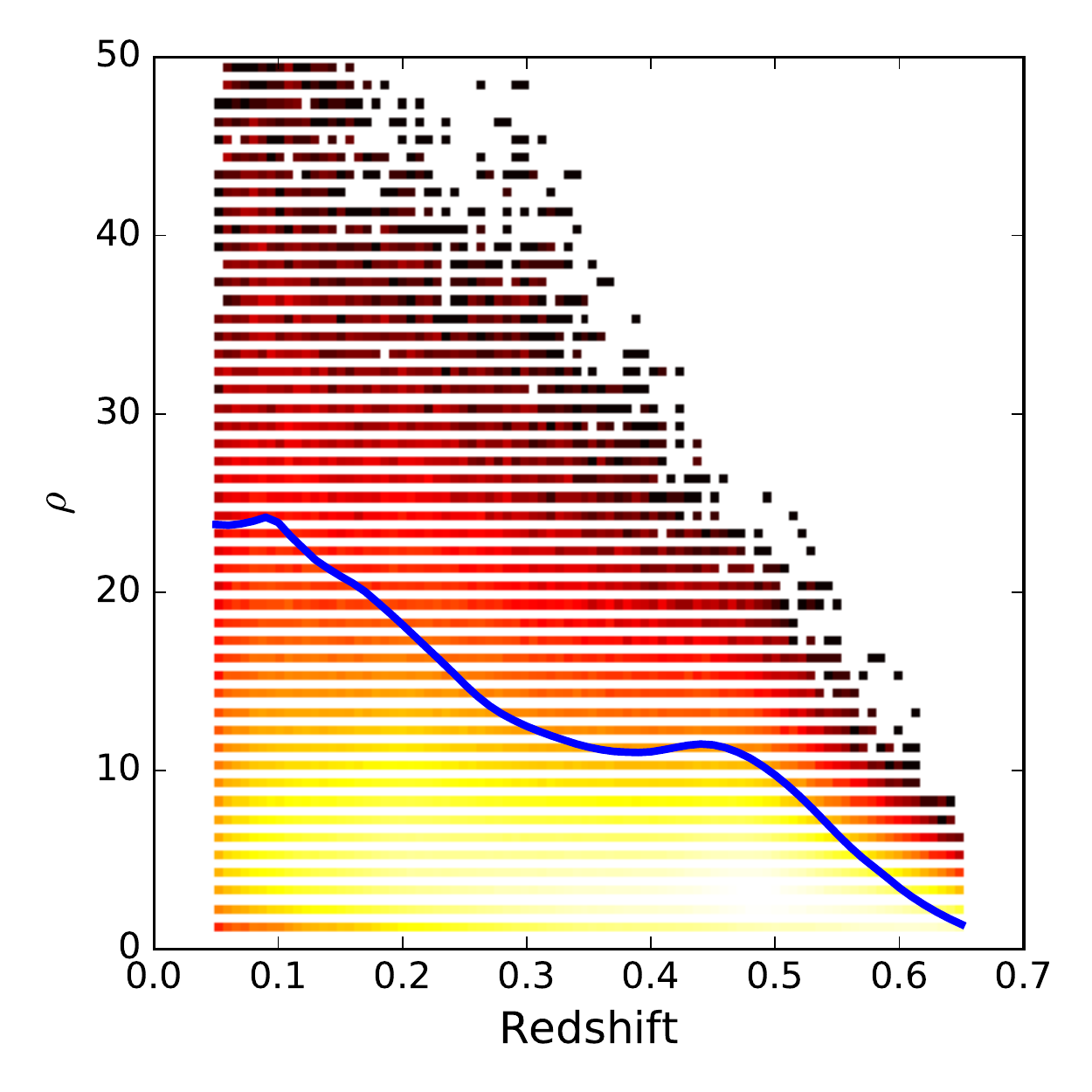}{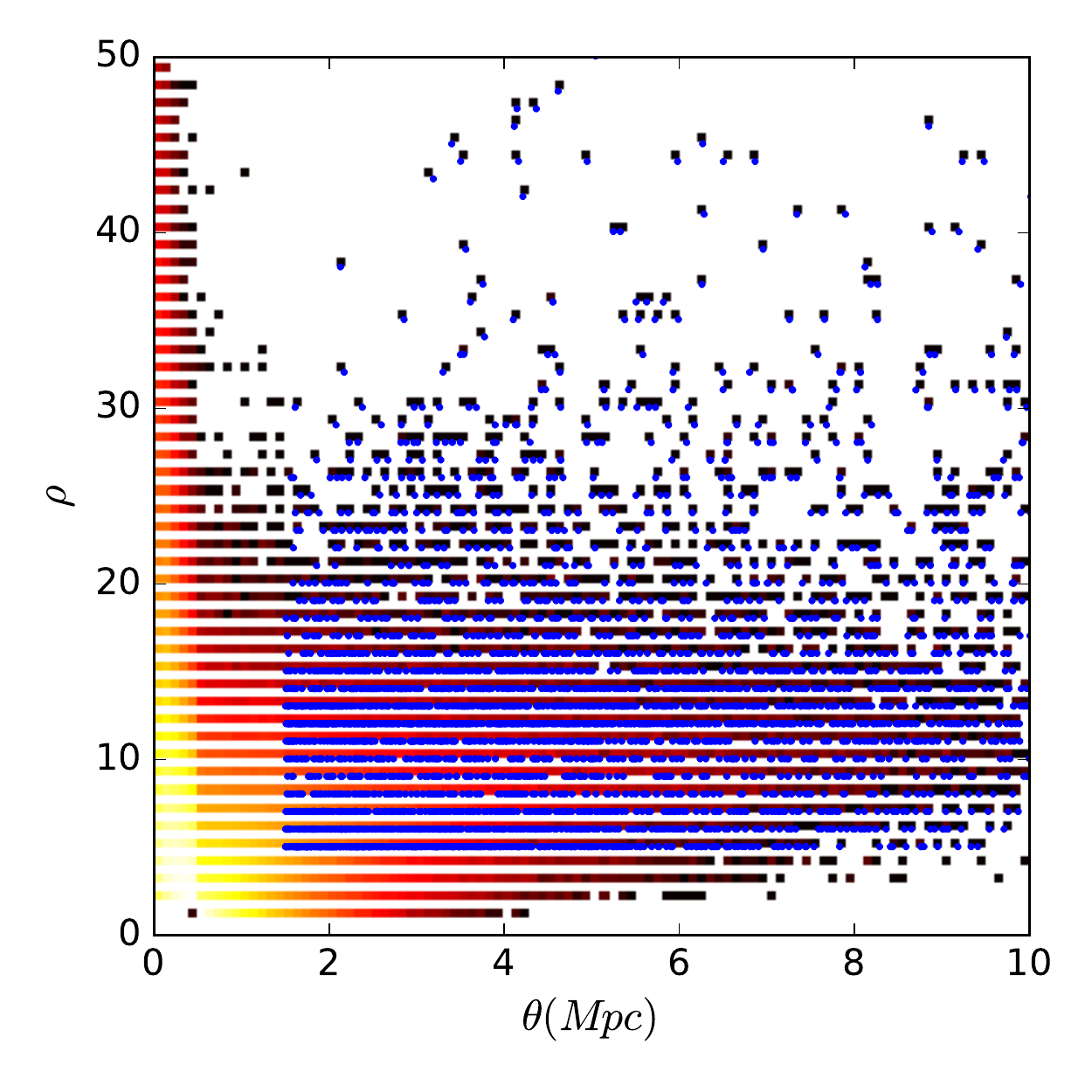}
\caption{Left: galaxy distribution on the plane of $\rho$ vs. $z$. The intensity map shows the galaxy density. The blue solid line shows $\rho = 5 \rho_\mathrm{field}$. Right: galaxy distribution on the plane of $\rho$ vs.  $\theta$. The cluster center candidates are marked as blue points.}
\label{fig-densityline}
\end{figure*}

\item Once the center candidates have been found, the galaxy members of a cluster are iteratively identified as the galaxies with $\theta<0.5$ Mpc linking to each other. The recognition of members is performed in a single step so it is much faster than the traditional friend of friend algorithm. 

\item The galaxy with the peak density is not always the brightest cluster galaxy (BCG). Usually the BCG is located near the densest region of the cluster and brighter than other members \citep{koe2007,hao2010}. We simply decide the BCG galaxy to be the brightest galaxy in $r$ band around the peak ($< 0.5$ Mpc). The BCG galaxy is considered as the final center. 

\item The redshift of the cluster is determined as the one of the BCG, denoted as $z_\mathrm{BCG}$. The members galaxies with $\left|z_\mathrm{phot}-z_\mathrm{BCG}\right| > 0.04*(1+z_\mathrm{BCG})$ are discarded. We count the numbers of remaining cluster members with distance $< 1$ Mpc ($N_\mathrm{1Mpc}$) and compute corresponding total $r$-band luminosity ($L_\mathrm{1Mpc}$ in unit of $L^{*}$). They are considered as proxies of the cluster richness. We require that the detected clusters have $N_\mathrm{1Mpc} \geqslant 12$.
\end{enumerate}

\begin{figure}[!ht]
\plotone{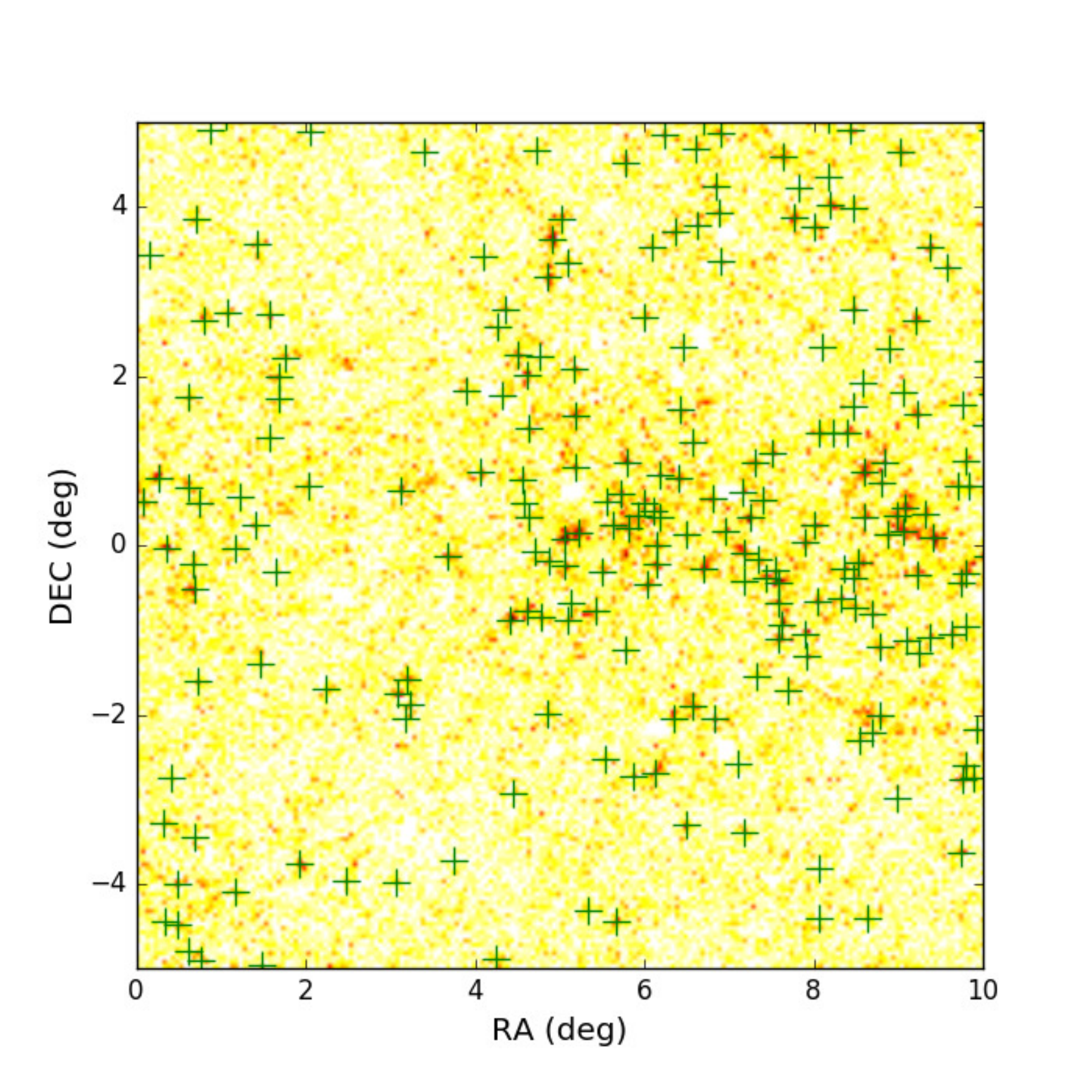}
\caption{Galaxy distribution within a redshift slice of $0.2 <z<0.3$ in a 100-deg$^2$ region. The intensity map shows the galaxy density. The detected cluster centers are marked with green pluses.}
\label{fig-slice}
\end{figure}

Figure \ref{fig-slice} shows an example of detected clusters on the sky. The intensity map shows the galaxy density within a redshift slice of $0.2 <z<0.3$. All the cluster centers, marked with plus, are located at the overdensity spots. There are some overdensity regions where no clusters are detected. Most of them are caused by the projection effect of foreground and background galaxies.

\subsection{Evaluating the algorithm}

Due to the projection effect and uncertainty of photo-z, our cluster finding algorithm might detect some false clusters. We evaluate the algorithm through performing a Monte Carlo simulation based on the real photometric data according to the methods used in \citet{wen2009} and \citet{hao2010}. We select an arbitrary region with the area of 500 deg$^2$. In this area, all galaxy members of identified clusters are taken away from the photometric redshift catalog. A random walk of 0--2 Mpc is assigned to each of the remaining background galaxies.The redshifts of the background galaxies are shuffled, where a galaxy is assigned with the  redshift of another randomly selected galaxy. This procedure makes a new random background and at the same time reserves original galaxy distribution on the sky caused by the projection effect. The identified clusters are finally put back to their original positions. A total of 10 simulations are performed and corresponding 10 mock catalogs are generated.

Our cluster finding procedure is implemented with the above 10 mock catalogs. The re-identified clusters are compared with original ones with a crossmatching separation of 1 Mpc and redshift tolerance of $|\Delta z| < 0.05$ following \citet{hao2010}. The clusters recognized in the mock catalog might contain false detections. Some of the original clusters might be submerged into the background. We define the completeness and purity to present the robustness of our detecting algorithm. The completeness is defined as the fraction of the original clusters that are found afresh from the mock catalog, while the purity is defined as the fraction of identified clusters found from the mock catalog that are also included in the original clusters. 

\begin{figure*}[!ht]
\plottwo{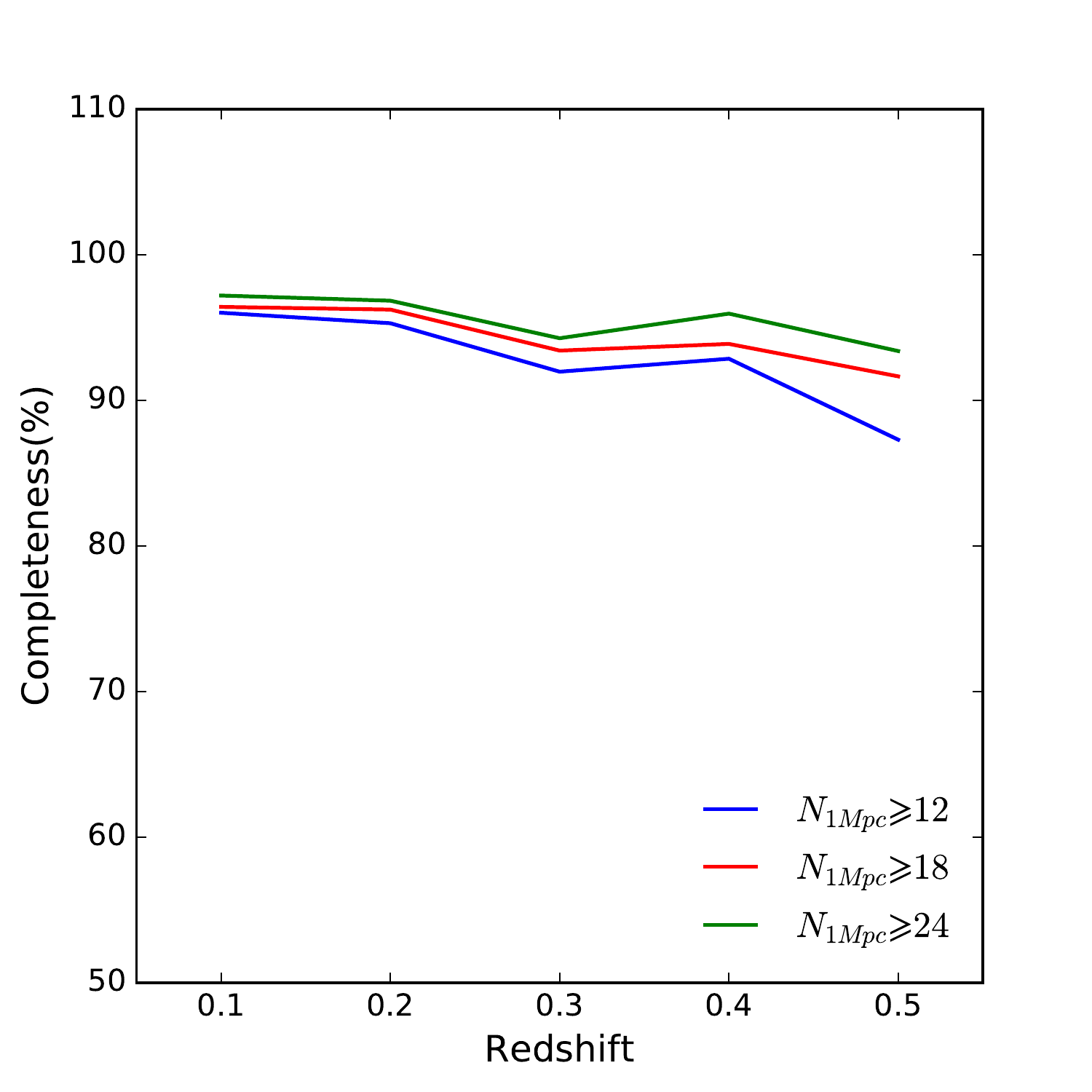}{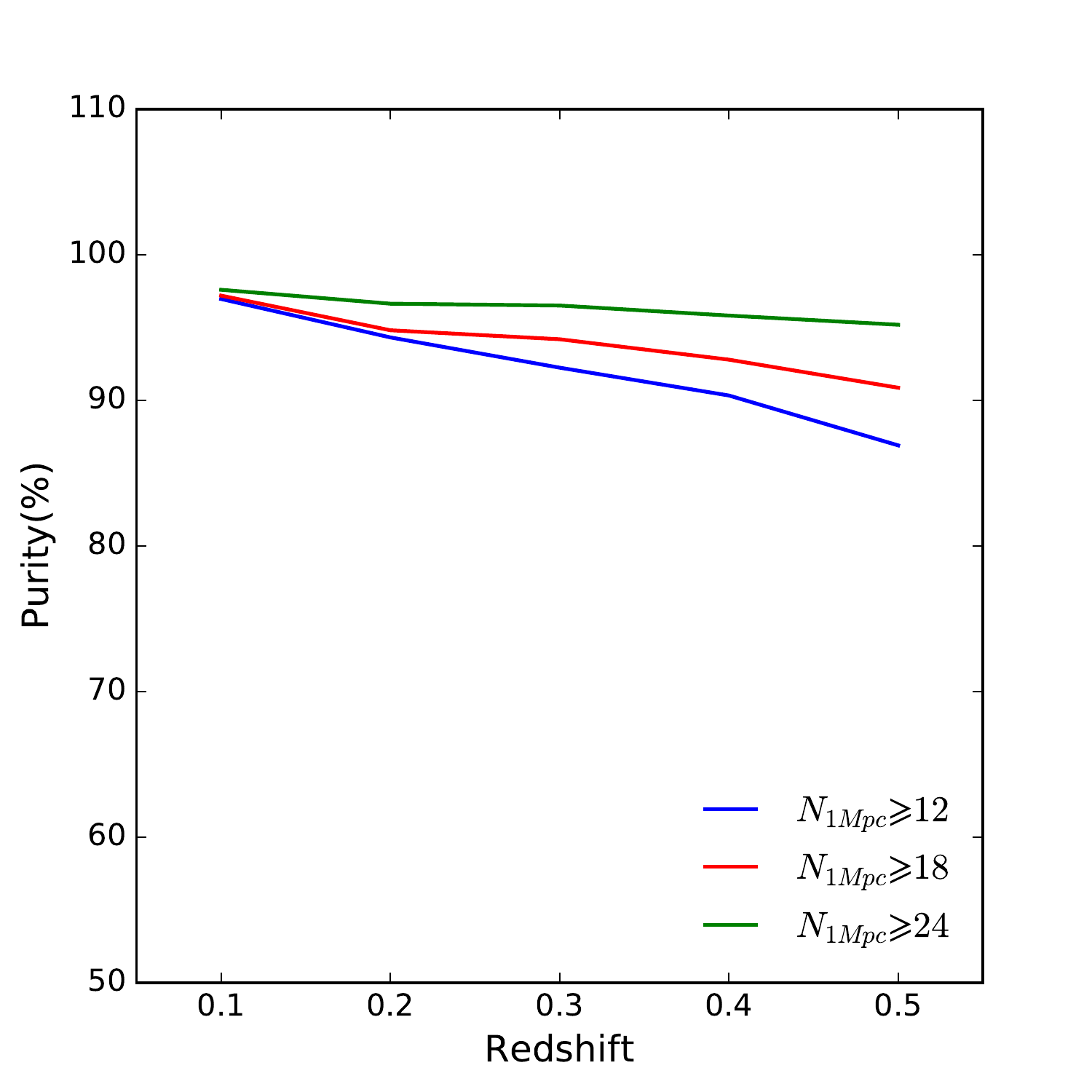}
\caption{Left: Average completenesses of clusters identified in the mock catalogs as function of redshift for different cluster richnesses ($N_\mathrm{1Mpc}$). Right: Average purities of clusters identified in the mock catalogs as function of redshift.}
\label{fig-mock}
\end{figure*}

Figure \ref{fig-mock} shows the average completenesses and purities of clusters identified in the mock catalogs as function of redshift for different cluster richnesses. As the redshift increases or $N_\mathrm{1Mpc}$ decreases, both completeness and purity decrease slightly. The overall completeness is about92\% (i.e. the fraction of original clusters that are re-identified). The completeness decreases from $97\%$ to $86\%$ in the redshift range of 0.1--0.5. For $N_\mathrm{1Mpc} \geqslant 24$, the completeness goes up to95.8\%. Most of missing clusters are either shifted to a new center with distance of $>1$ Mpc away from the original locations or assigned with a different redshift ($|\Delta z| > 0.05$). There are a fraction of the missing clusters (about $27.5\%$) with the calculated richness of $N_\mathrm{1Mpc} < 12$. The purity is equivalent to the false detection rate. The overall false detection rate is about8.9\%. The false rate drops to3.6\% for $N_\mathrm{1Mpc} \geqslant 24$. Most of the false detections have relatively low richnesses.

\subsection{Richness and mass estimations} \label{sec:richness}
Richness is one of the primary parameters of galaxy clusters, which traces the cluster mass. The detection of galaxy clusters is affected by the selection effect of the photometric data. This kind of selection effect causes an underestimation of richness for clusters at relatively high redshift. \citet{wen2015} defined a richness estimator,  $R_{L*,500}$, which is calibrated by clusters with X-ray and SZ measurements. This richness estimator is independent on redshift. Following \citet{wen2018}, we use $R_{L*,500}$ to calibrate our richness estimator, $R_{L^*}$, which is defined as:
\begin{equation}
R_{L^*}=a*L^b_\mathrm{1Mpc}*(1+z)^c, 
\label{equ_1}
\end{equation}
where $L_\mathrm{1Mpc}$ is the total $r$-band luminosity of cluster members within a distance of 1 Mpc from the BCG, $a$ is a linear coefficient, and $b$ and $c$ are the power indices for luminosities and the correction of redshift dependency, respectively. We get a sample of202 clusters through matching our clusters to those with X-ray and SZ measurements in \citet{wen2015}. Based on $R_{L*,500}$ in \citet{wen2015} and our measured $L_\mathrm{1Mpc}$ of these clusters, we derive best-fit richness estimator as below:
\begin{equation}
R_{L^*}=0.69*L^{1.32}_\mathrm{1Mpc}*(1+z)^{2.91}.
\label{equ_2}
\end{equation}

\begin{figure*}[!ht]
\plottwo{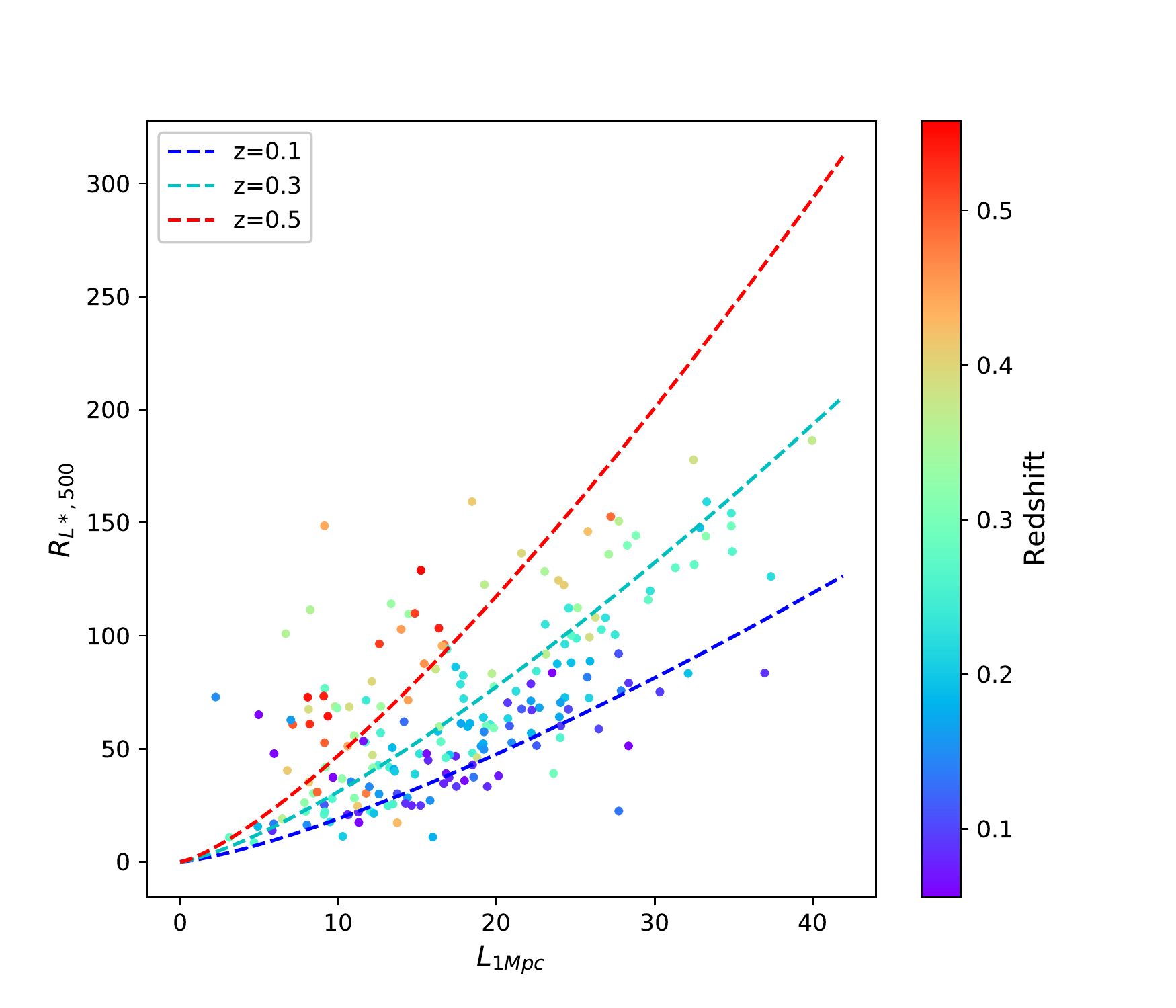}{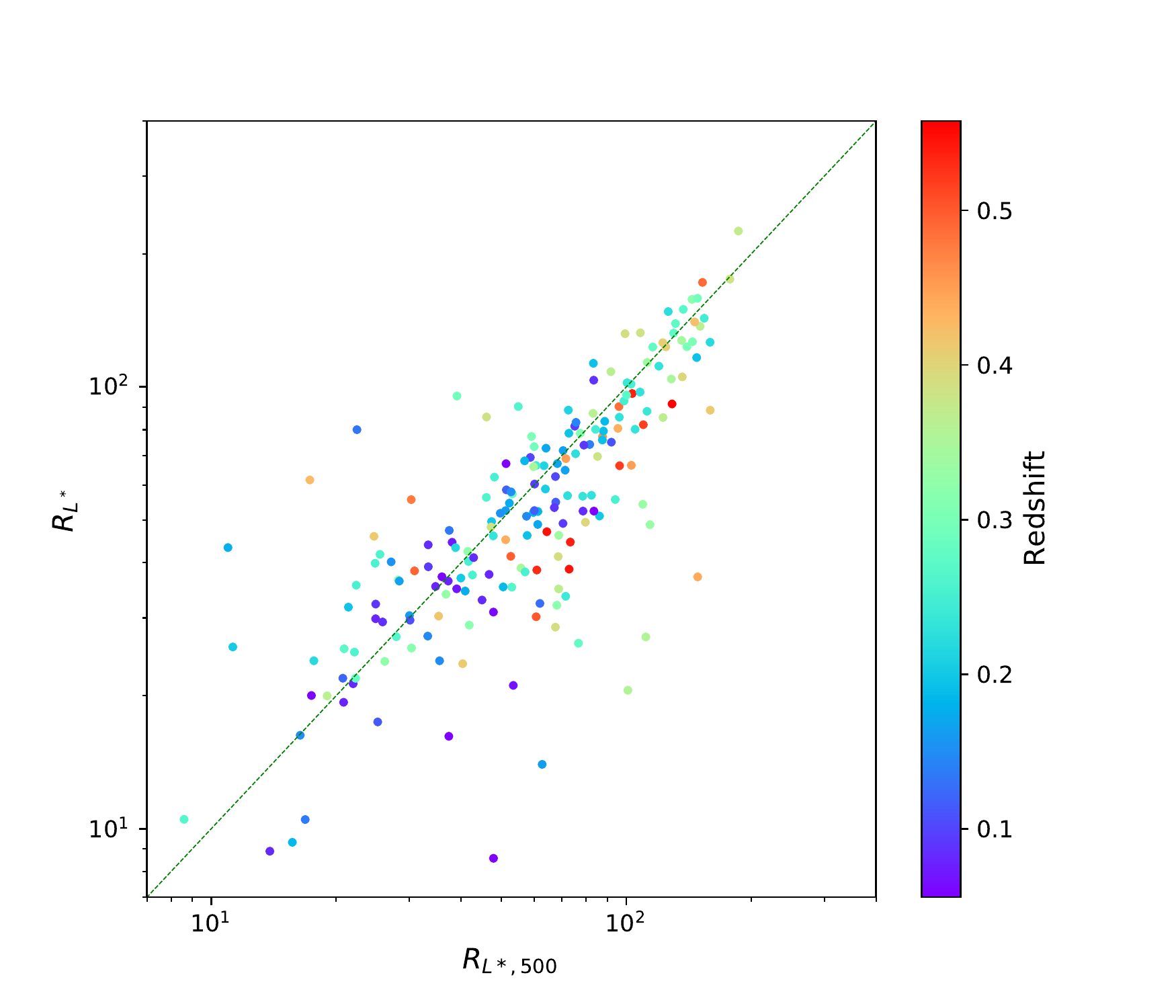}
\caption{Left: The relation between $R_{L*,500}$ and $L_\mathrm{1Mpc}$. The fitted relations at $z=0.1$, 0.3, and 0.5 as shown in Equation (\ref{equ_2}) are overplotted.  Right: The comparison between $R_{L*,500}$ in \citet{wen2015} and our richness $R_{L^*}$. The points are colorred by the spectroscopic redshift.}
\label{fig-richness}
\end{figure*}

The left panel of Figure \ref{fig-richness} shows the relation between $R_{L*,500}$ and $L_\mathrm{1Mpc}$ colorred by redshift. The fitted relations between  $R_{L*,500}$ and $L_\mathrm{1Mpc}$ as shown in Equation (\ref{equ_2}) at $z=0.1$, 0.3, and 0.5 are overplotted in this figure.  The right panel of Figure \ref{fig-richness} shows the comparison between $R_{L*,500}$ in \citet{wen2015} and our richness $R_{L^*}$. The scatter is about 17.5 after removing the outliers with a $3\sigma$-clipping algorithm. There is no significant redshift dependency of the richness difference. The cluster mass is estimated using the scaling relation between the mass and richness as shown in Equation (17) of \citet{wen2015}, which is expressed as $\log M_{500} = (1.08 \pm 0.02)\log R_{L*} - (1.37 \pm 0.02)$.

\section{The cluster catalog} \label{sec:catalog}
\subsection{Catalog description and examples}
\begin{figure*}[!ht]
\subfigure{\includegraphics[width=0.33\textwidth]{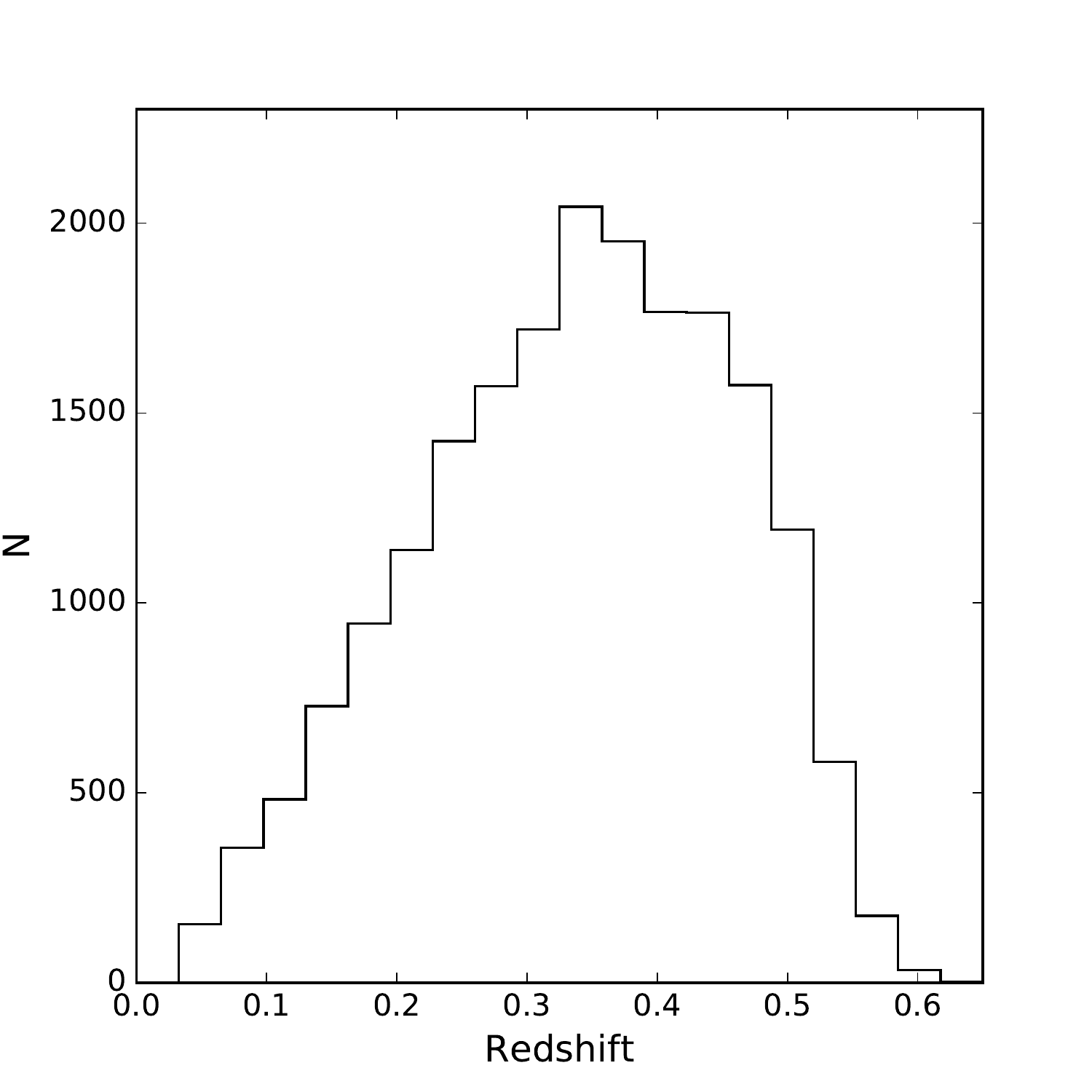}}
\subfigure{\includegraphics[width=0.33\textwidth]{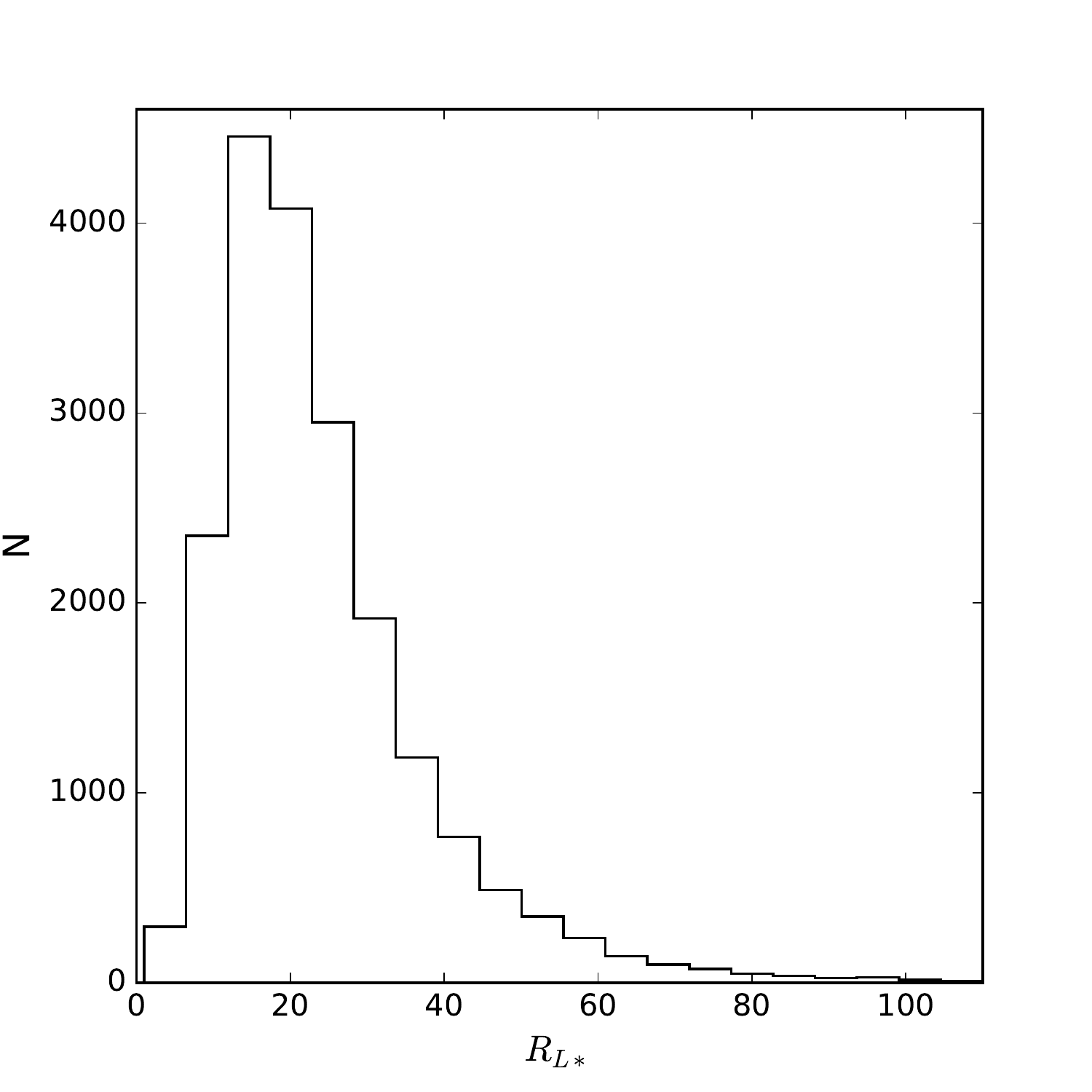}}
\subfigure{\includegraphics[width=0.33\textwidth]{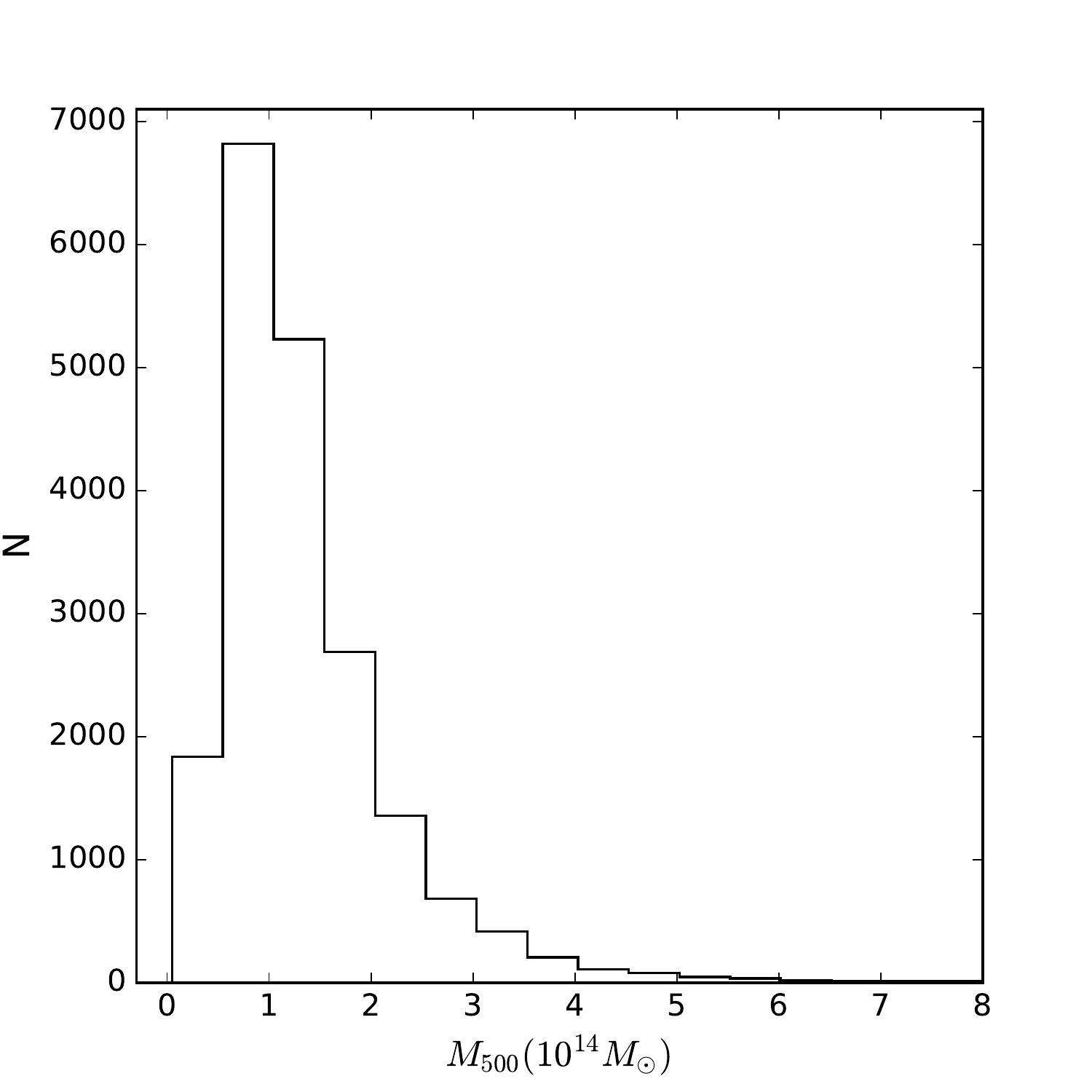}}
\caption{Left: redshift distribution of galaxy clusters in our catalog. Middle: richness distribution. Right: mass distribution.}
\label{fig-distribution}
\end{figure*}

Our cluster catalogs contains19,610 clusters detected from the photometric redshift catalog based on SCUSS, SDSS, and unWISE. The catalog is accessible online\footnote{\url{http://batc.bao.ac.cn/~zouhu/doku.php?id=projects:scuss_clusters:start}}.  The redshift coverage is between 0.05 and 0.65. Figure \ref{fig-distribution} shows the distributions of the redshift, richness, and mass for the detected clusters. The median redshift, richness, and mass are 0.35, 21.7, and $1.2\times10^{14} M_\sun$, respectively.  Table \ref{tab-catalog} lists the contents included in the catalog. The cluster centers are located at the positions of BCGs. The photometric redshifts and spectroscopic redshifts if available for the clusters are provided. The richness and mass of the clusters are estimated from the $r$-band luminosity $L_\mathrm{1Mpc}$.

\begin{table*}
\label{tab:catalog}
\centering
\scriptsize
\caption{Catalog contents of our galaxy cluster catalog}\label{tab-catalog}
\begin{tabular}{ccl}
\tableline
\tableline
Column   & Unit & Description\\
\tableline
NAME & \nodata &Cluster order number  \\
RA & degree & R.A. in J2000 of the BCG \\
DEC& degree & Declination in J2000 of the BCG \\
Z\_PHOTO & \nodata & Photometric redshift of the BCG \\
Z\_SPEC & \nodata & Spectroscopic redshift of the BCG if existing (or else -10)  \\
Z\_PHOTO\_MEDIAN & \nodata & Median photometric redshift of member galaxies \\
N\_1MPC & \nodata & Number of member galaxies around BCG galaxy within a radius of 1 Mpc\\
L\_1MPC & $L^{*}$ & The integrated luminosity of the member galaxies around BCG galaxy within a radius of 1 Mpc \\
RICHNESS & \nodata & Estimated richness calibrated using the catalog of \citet{wen2012} \\
M\_500 & $10^{14} M_\sun$ & Estimated cluster mass within the characteristic radius\tablenotemark{a}\\

\tableline
\end{tabular}
\tablenotetext{a}{The characteristic radius is defined as the radius within which the mean mass density is 500 times that of the critical cosmic mass.}
\end{table*}
		
\begin{figure*}[!ht]
		\subfigure{\includegraphics[width=0.33\textwidth]{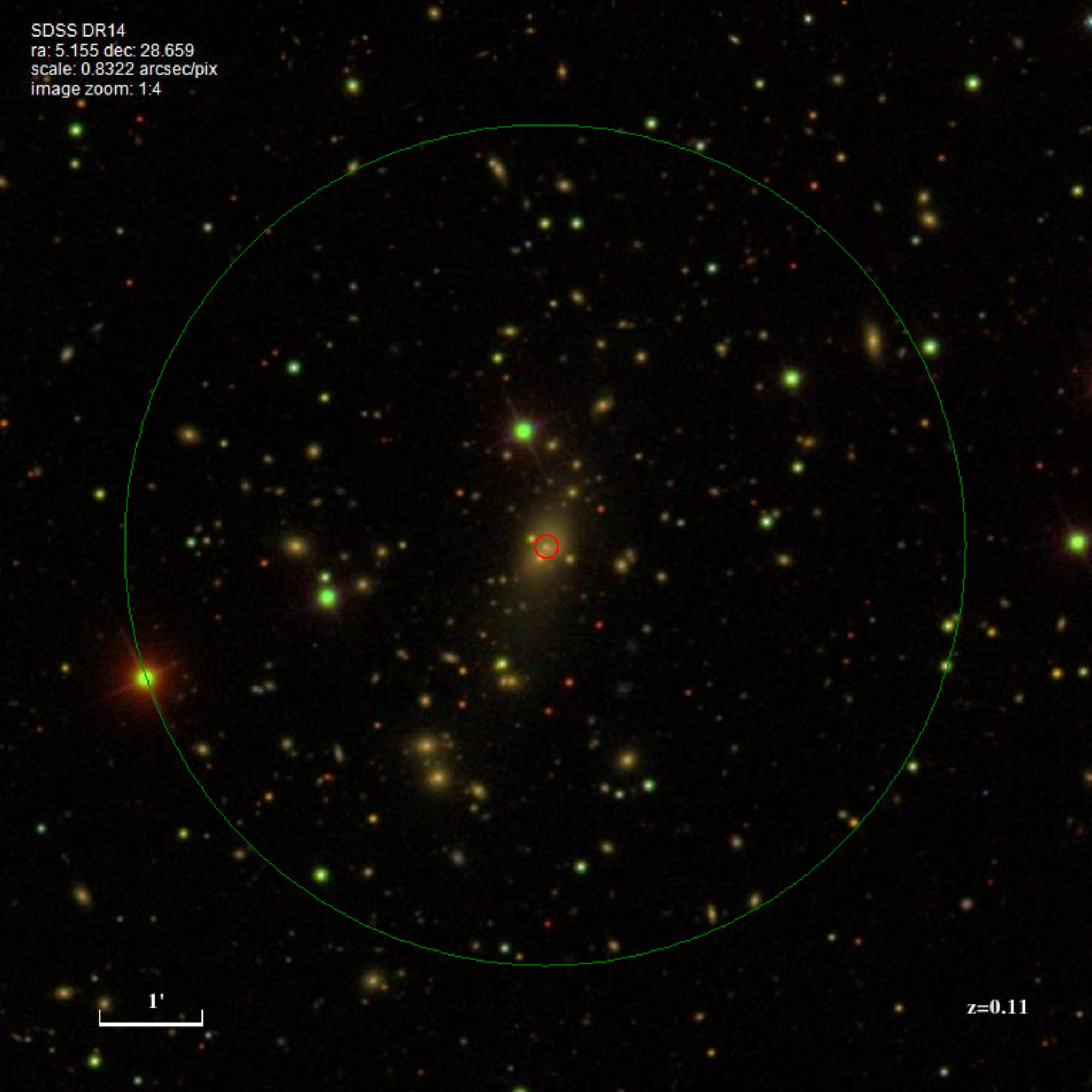}}
		\subfigure{\includegraphics[width=0.33\textwidth]{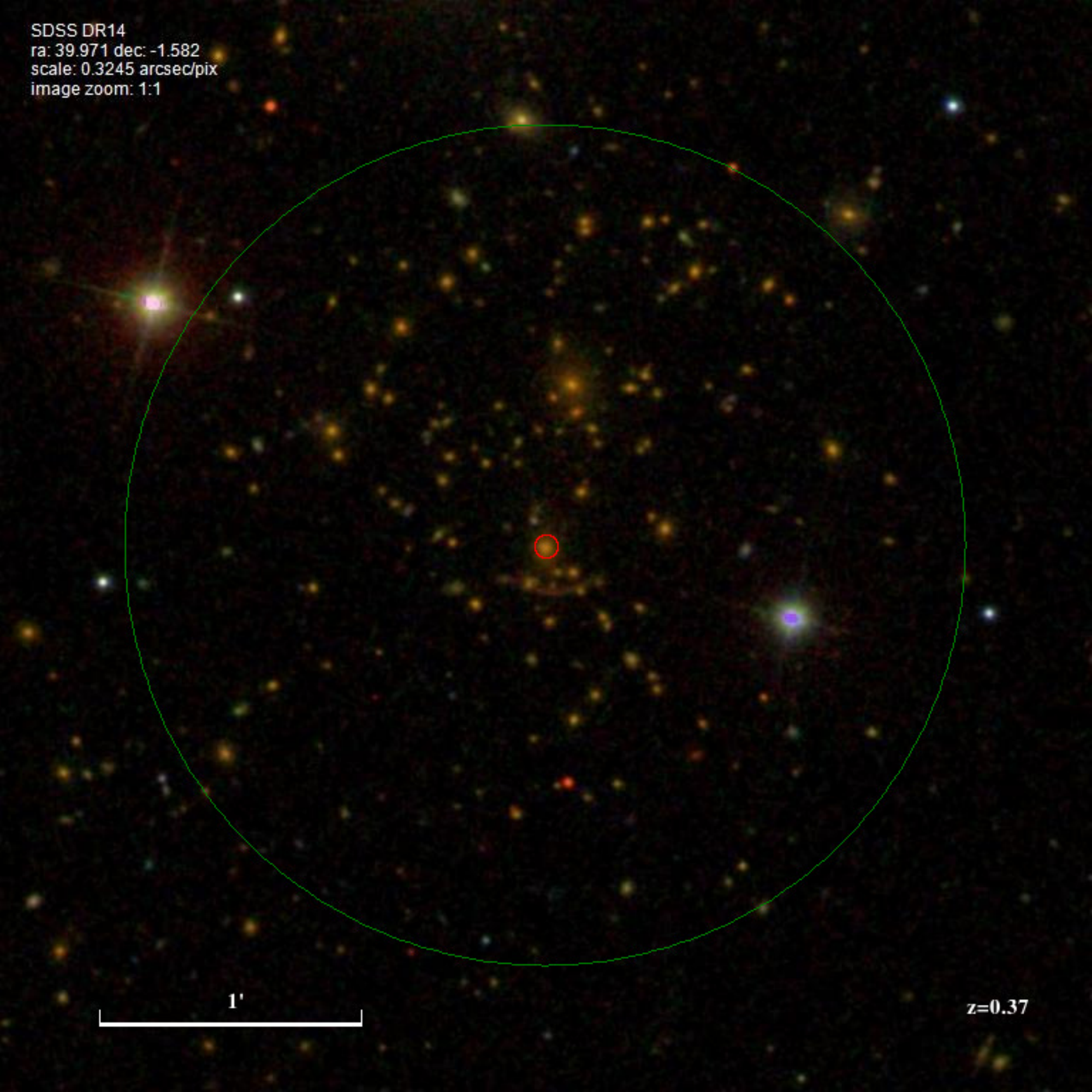}}
		\subfigure{\includegraphics[width=0.33\textwidth]{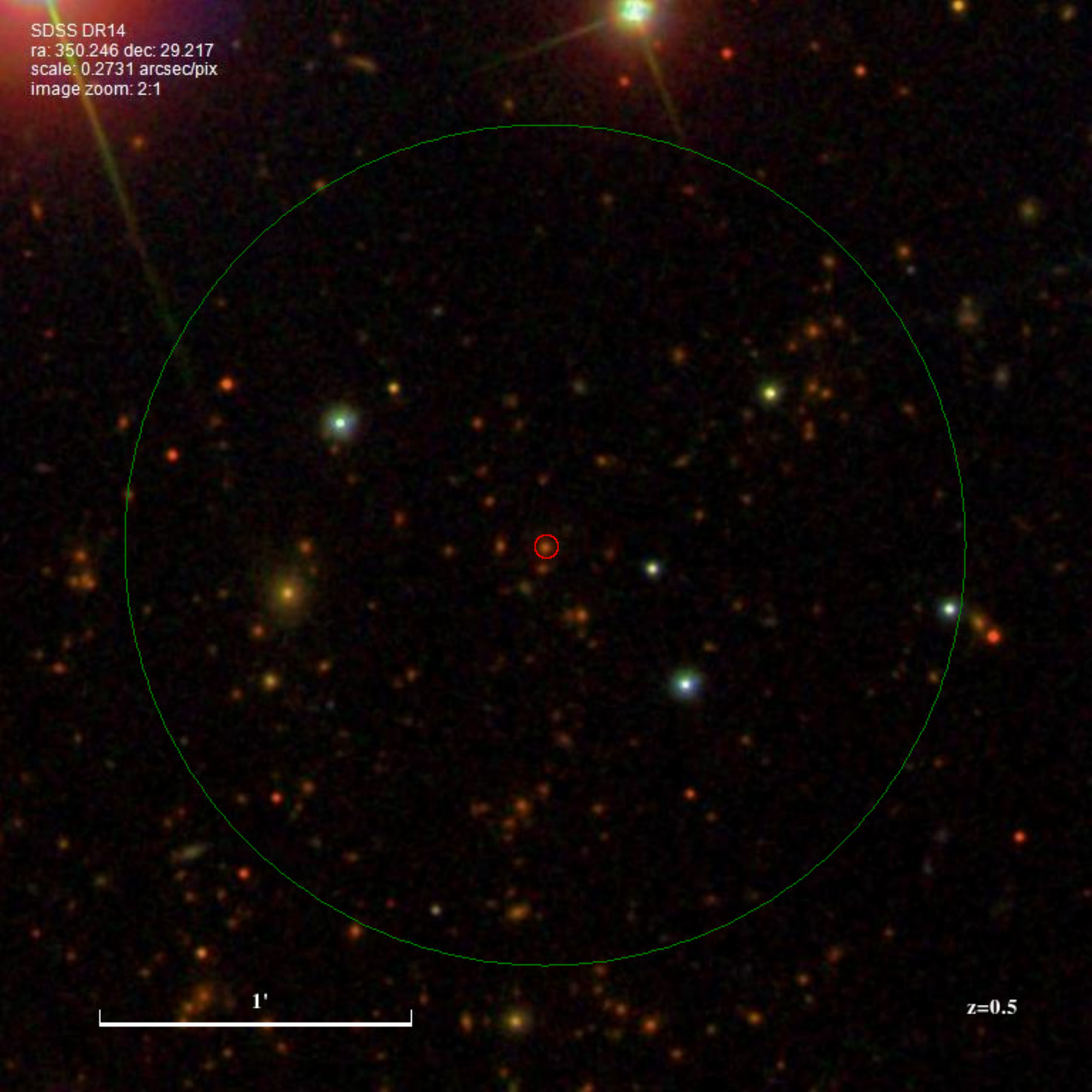}}		
		\caption{SDSS color images of three rich clusters detected at different redshifts of $z=0.11$ (left), 0.37 (middle), and 0.5 (right). The BCGs are marked with red circles. The bigger green circle demonstrates the distance of 0.5 Mpc from the center.}
		\label{fig-sample}
		\end{figure*}

\begin{figure*}[!ht]
		\plottwo{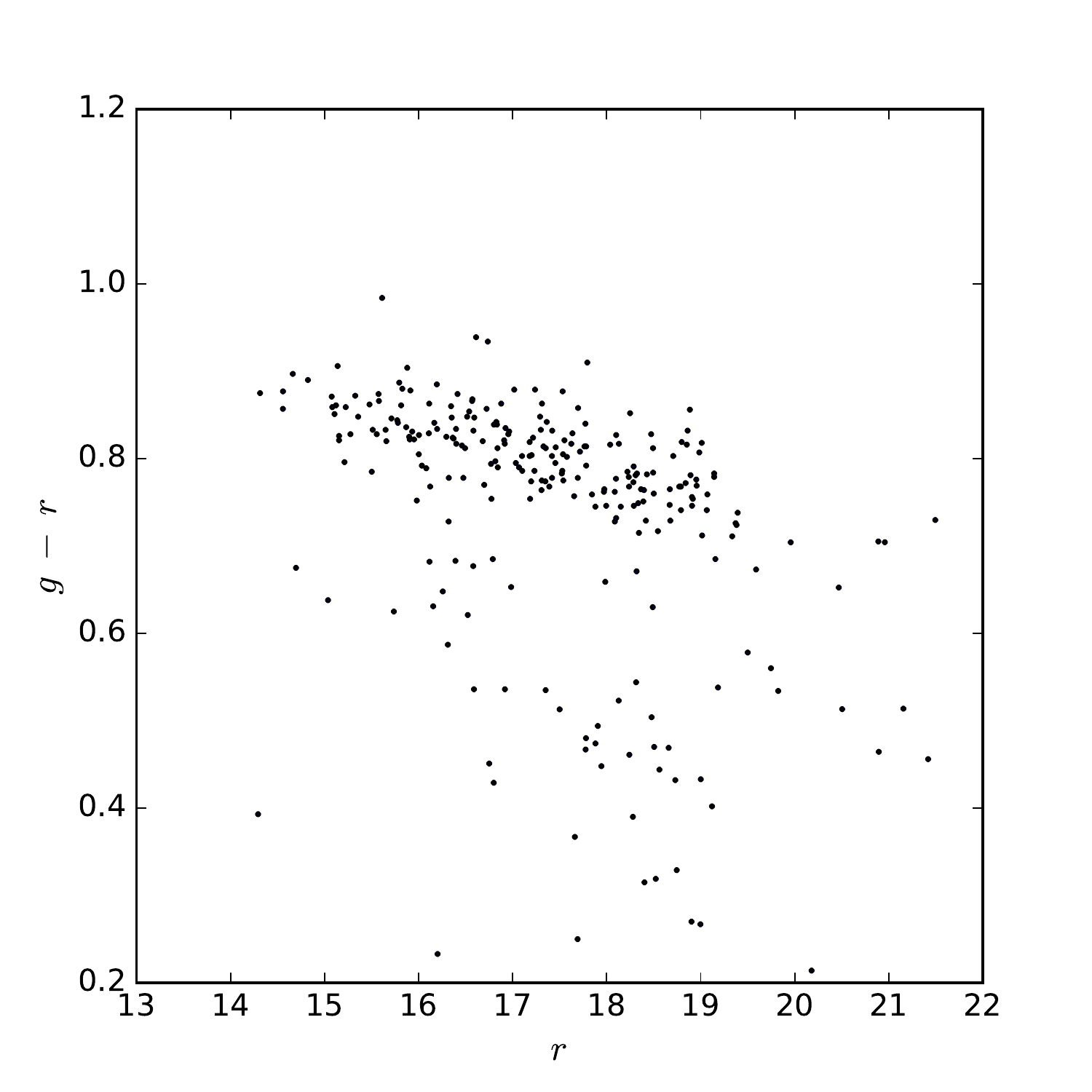}{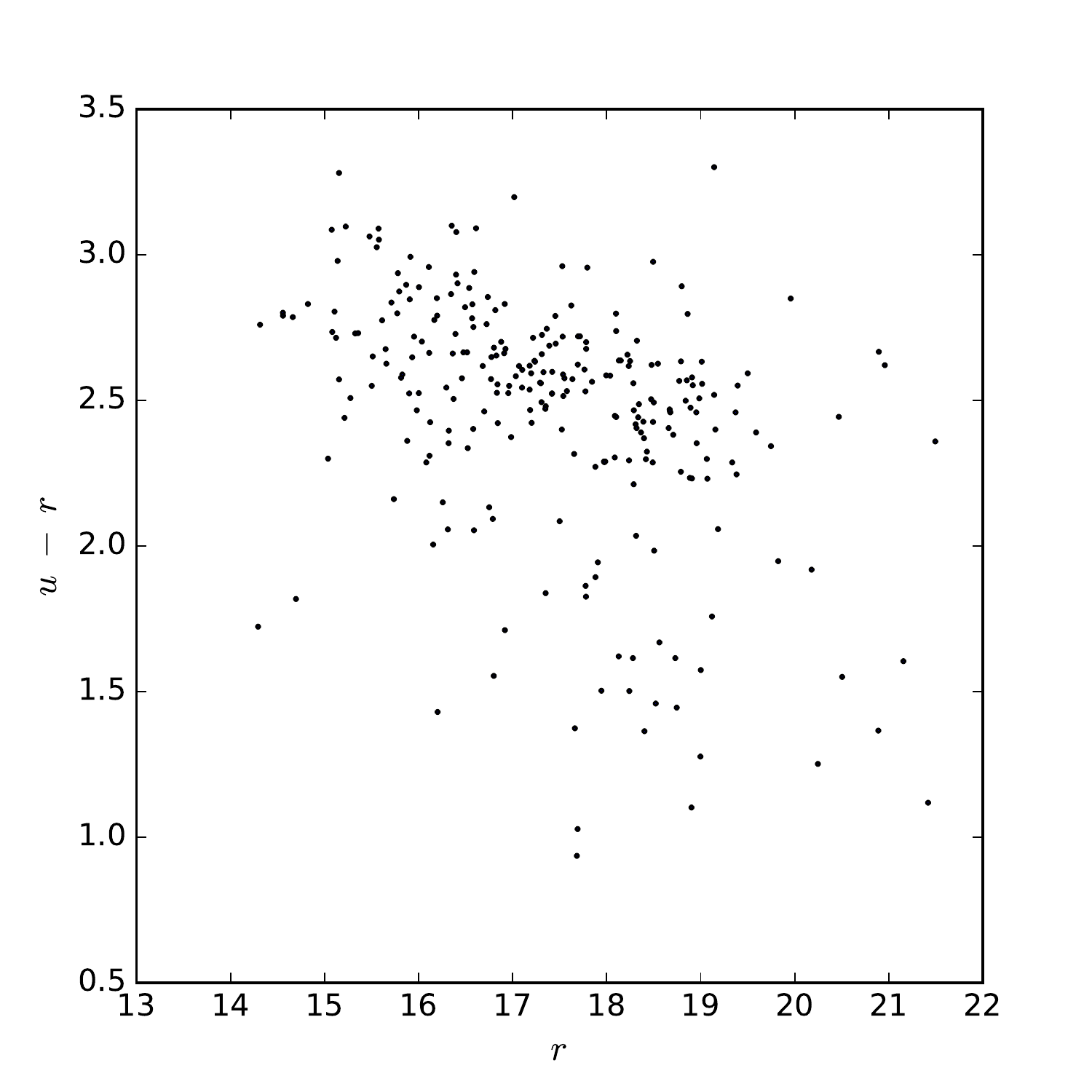}
		\caption{Left: Color-magnitude diagram of $g-r$ vs. $r$ for a rich cluster. Right: Color-magnitude diagram of $u-r$ vs. $r$ for the same cluster.}
		\label{fig-cmd}
		\end{figure*}

Figure \ref{fig-sample} exhibits SDSS color images of three rich clusters at different redshifts. The color images are retrieved from the SDSS webpage\footnote{\url{http://skyserver.sdss.org/dr14/en/tools/chart/image.aspx}}. The clusters are centered on BCGs. From these images, we can see that a considerable number of elliptical (red) galaxies are concentrated. Figure \ref{fig-cmd} shows an example of color-magnitude diagrams for member galaxies of one arbitrarily selected rich clusters. Both plots of $g-r$ vs. $r$ and $u-r$ vs. $r$ show the sequence of red galaxies. The $u-r$ color presents a larger scatter for the red sequence, because the $u$-band flux is more insensitive to the star formation and dust extinction.
		
\subsection{Photometric redshift uncertainty of clusters}
\begin{figure*}[!ht]
\plottwo{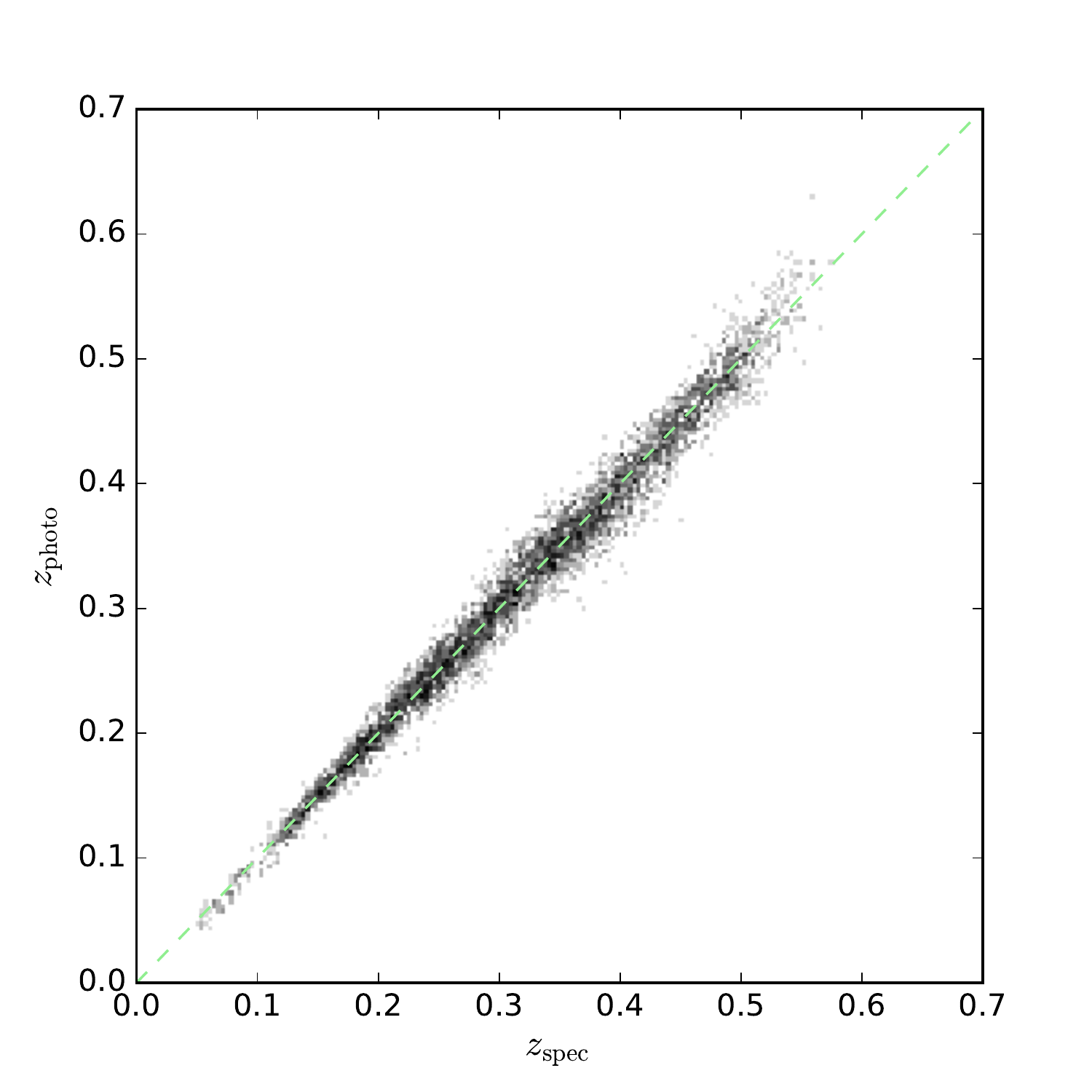}{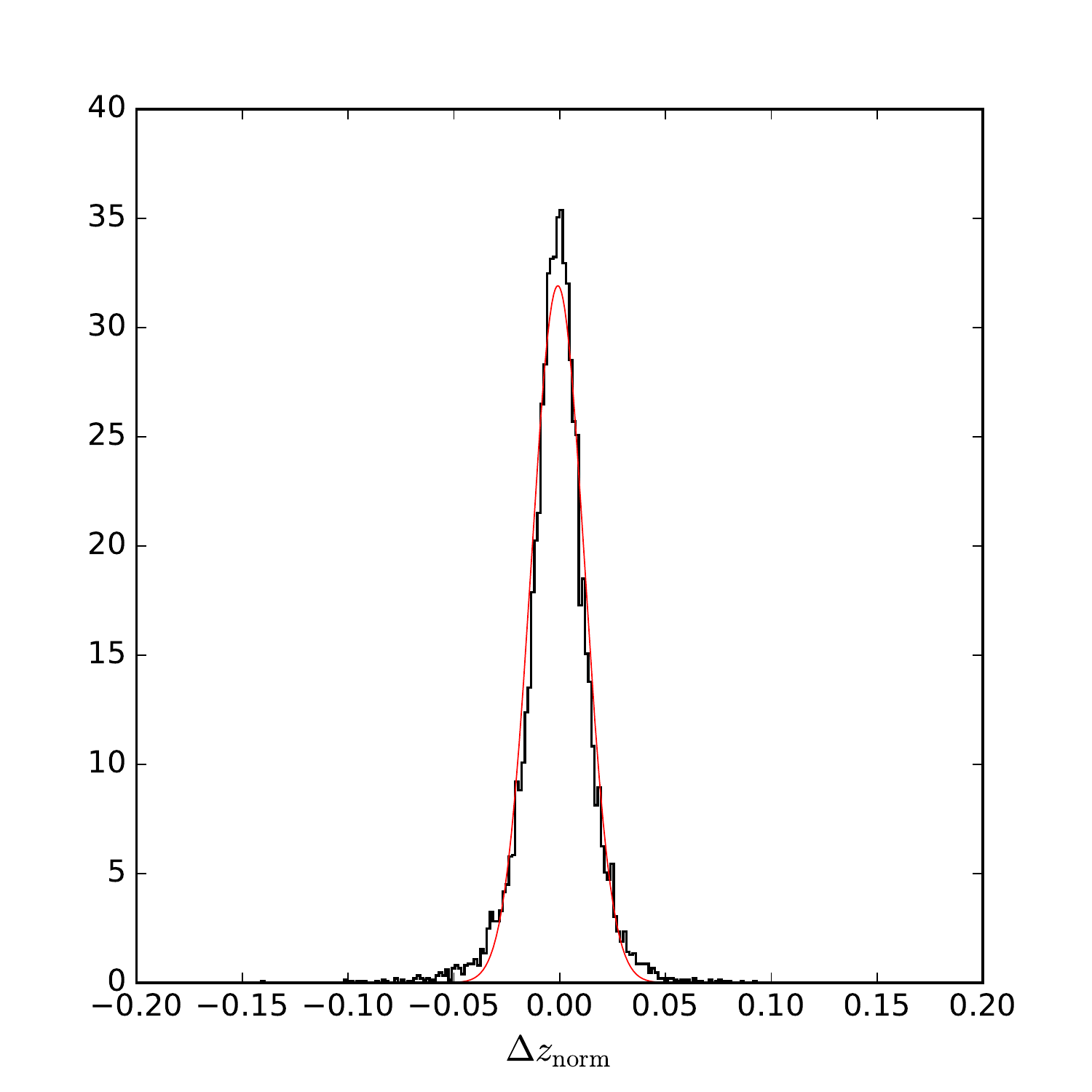}
\caption{Left:  $z_\mathrm{phot}$ vs. $z_\mathrm{spec}$ for BCGs. The green dashed lines shows $z_\mathrm{spec}=z_\mathrm{phot}$. The grey scale shows the galaxy density. Right:  normalized distribution of $\Delta z_\mathrm{norm}$. The red solid line shows the curve giving the best-fit Gaussian function with $\sigma =0.0125$ and  $\mu=-0.000853$.}
\label{fig-photoz}
\end{figure*}

The redshift of the BCG is taken as the redshift of a cluster in our catalog since BCGs have better photo-z estimations than other members. Some of the BCGs have spectroscopic redshift measurements. We present the redshift uncertainty of our clusters by comparing with the spectroscopic redshifts with photometric redshifts of the BCGs in Figure \ref{fig-photoz}. The left panel of this figure shows the $z_\mathrm{spec}$ vs. $z_\mathrm{phot}$. The right panel displays the distribution of $\Delta z_\mathrm{norm}$, which is fitted with a Gaussian function. The expectation of  $\Delta z_{norm}$ is $\mu=-0.000853$ and the standard deviation is $\sigma \Delta z_\mathrm{norm} =0.0125$. The photometric redshift uncertainty of our clusters is smaller than the general uncertainty of the photo-z catalog ($\sim$ 0.02). By contrast, the redshift uncertainties of other cluster catalogs are calculated as 0.015 for WHL2012, 0.015 for GMBCG, and 0.009 for maxBCG and corresponding biases are 0.0013, 0.0018, and 0.003 for these catalogs, respectively. The maxBCG catalog contains clusters at relatively low redshift ($0.1<z<0.3$), so the redshift uncertainty is smaller.   

\subsection{Comparison with other catalogs} \label{sec:comp}
As mentioned previously, there have been a number of galaxy cluster catalogs that are based on SDSS data. The discrepancy between these catalogs comes from several aspects, including the photometric data, photometric redshift estimation, and cluster detecting algorithm, etc. Our detection of galaxy clusters are based on three imaging surveys. The application of unWISE imposes the most significant limit to the number of galaxies having good photometric redshifts due to the shallower depth and lower resolution relative to the optical data. However, the SCUSS $u$ and unWISE $W1W2$ help to improve the photo-z quality. In the following sections, we mainly make some comparisons with the Abell and WHL2012 catalogs.

\subsubsection{Comparison with the Abell Clusters}
The Abell clusters are the most famous and clear on the sky. The Abell catalog is obtained from CDS\footnote{\url{http://vizier.u-strasbg.fr/viz-bin/VizieR}}, including a total of 2,712 clusters compiled by \citep{abe1989}. There are 465 clusters located in the SCUSS footprint.  Most of the clusters have no redshift information.  Among the clusters with redshifts, 19 clusters have $z < 0.05$ and are eliminated in our comparison due to the redshift limit. The remaining 446 Abell clusters are used to match with our catalog using our photometric redshifts and a separation of 2 Mpc. There are 422 matched clusters, accounting for about 95\% of the total sample. Abell 64 and Abell 65 clusters are assigned to the same cluster of ID = 2309 in our catalog. Abell 2571 and Abell 2573 are assigned to the same cluster of ID = 17127 in our catalog. These cluster pairs are close to each other, so that they are considered as one single cluster by our cluster detection algorithm. Among the 24 un-matched Abell clusters, there are 16 clusters locating at either the edges of the footprint or the regions where galaxies have no photometric redshifts due to invalid photometry. The rest 8 Abell clusters are lost due to relatively low $\rho$ or $\theta$ values. If not counting the above 16 clusters, we get a matching rate of about 98\%.

\subsubsection{Comparison with the WHL2012 clusters}
\citet{wen2012} identified 132,684 galaxy clusters at $0.05<z<0.8$ based on the photometric redshifts from SDSS-III. These clusters were obtained by applying a friend of friend algorithm to detecting the overdensity features. The WHL2012 catalog is the largest catalog, which used only SDSS data and similar detecting method to this work. It was stated in \citet{wen2012} that the clusters at $z>0.42$ were less complete and had a biased smaller richness due to incompleteness of member galaxies. \citet{wen2015} updated the WHL2012 cluster catalog with a new richness estimator that was calibrated the optical mass proxy using the X-ray and SZ measurements.  We use the updated catalog for comparison. 

\begin{figure*}[!ht]
		\subfigure{\includegraphics[width=0.33\textwidth]{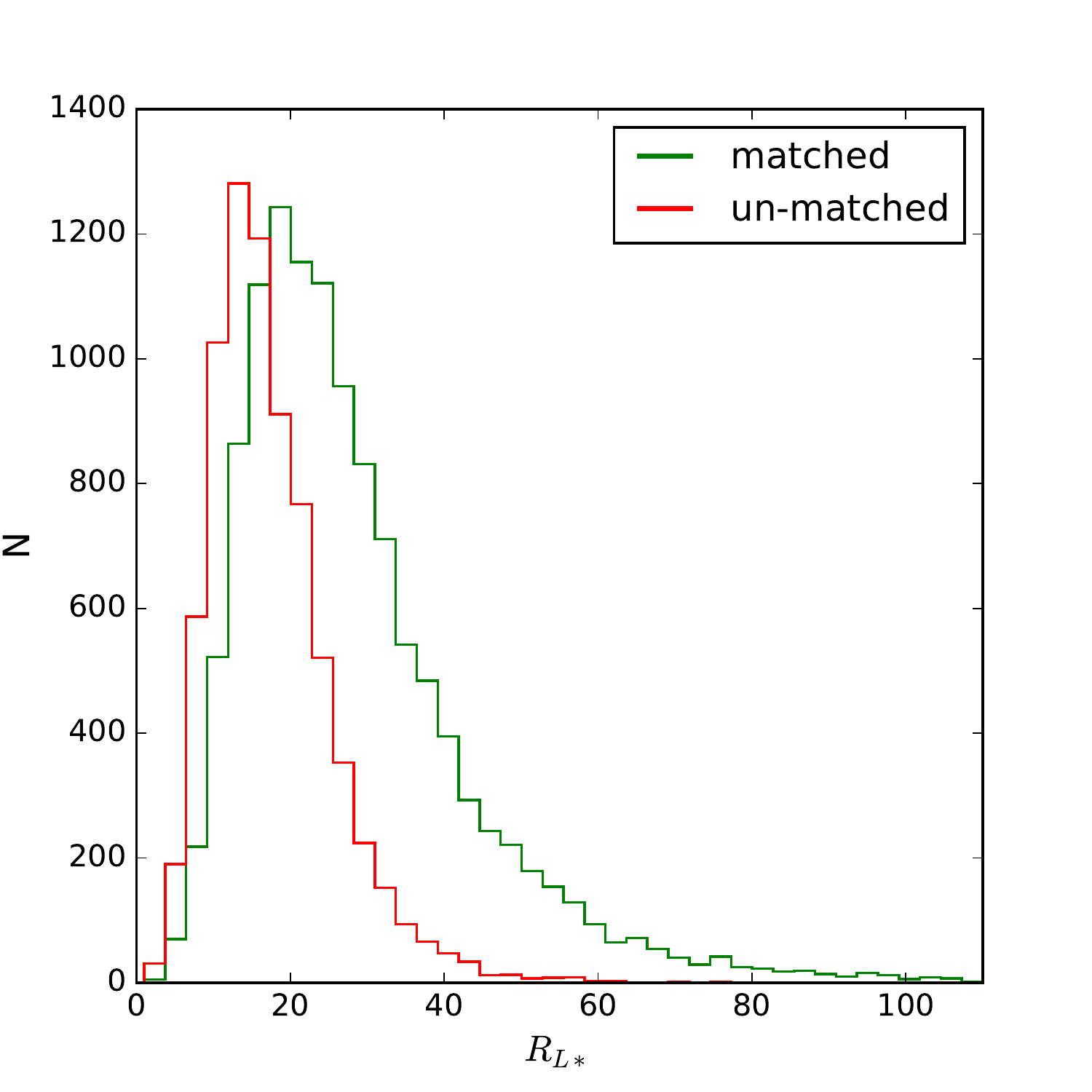}}
		\subfigure{\includegraphics[width=0.33\textwidth]{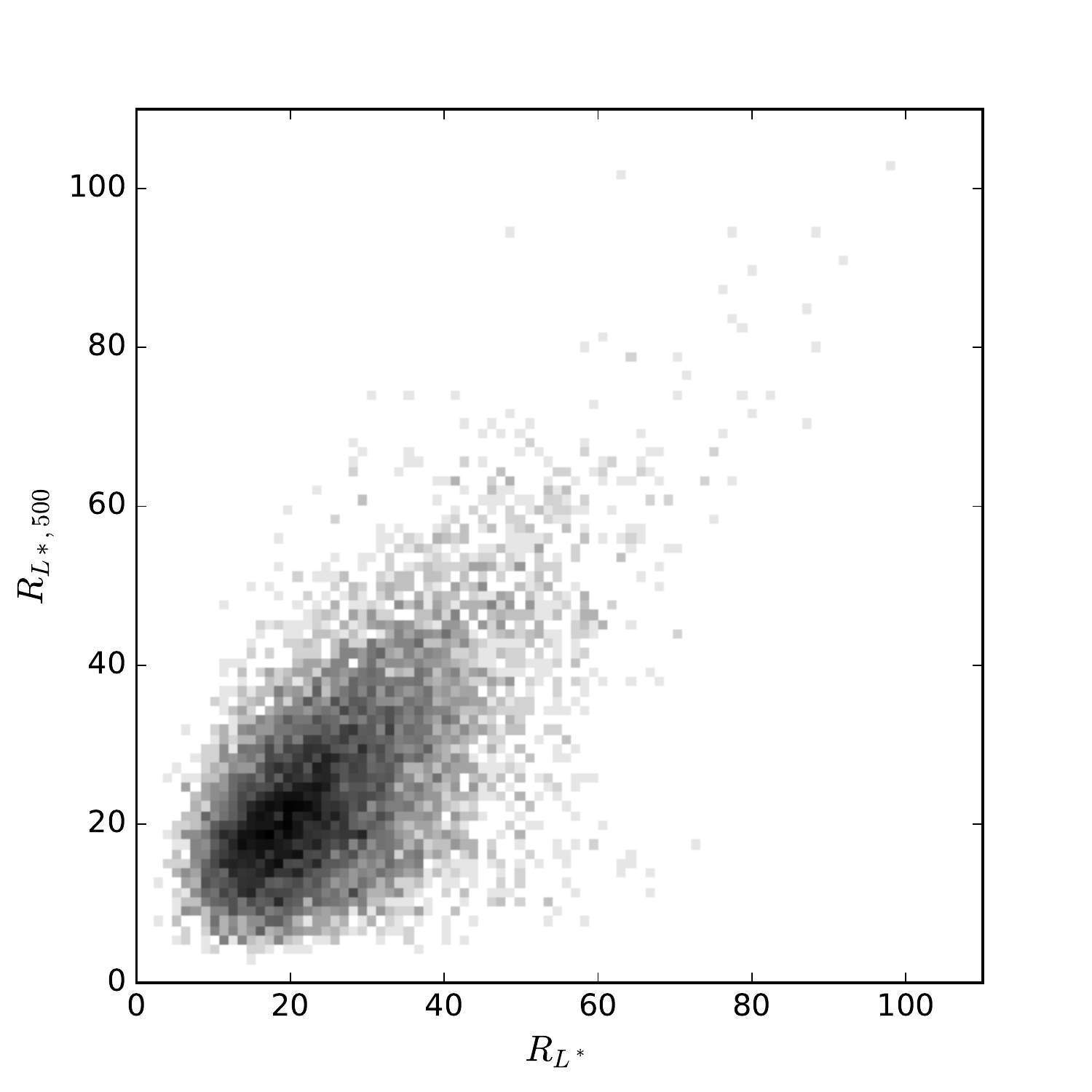}}
		\subfigure{\includegraphics[width=0.33\textwidth]{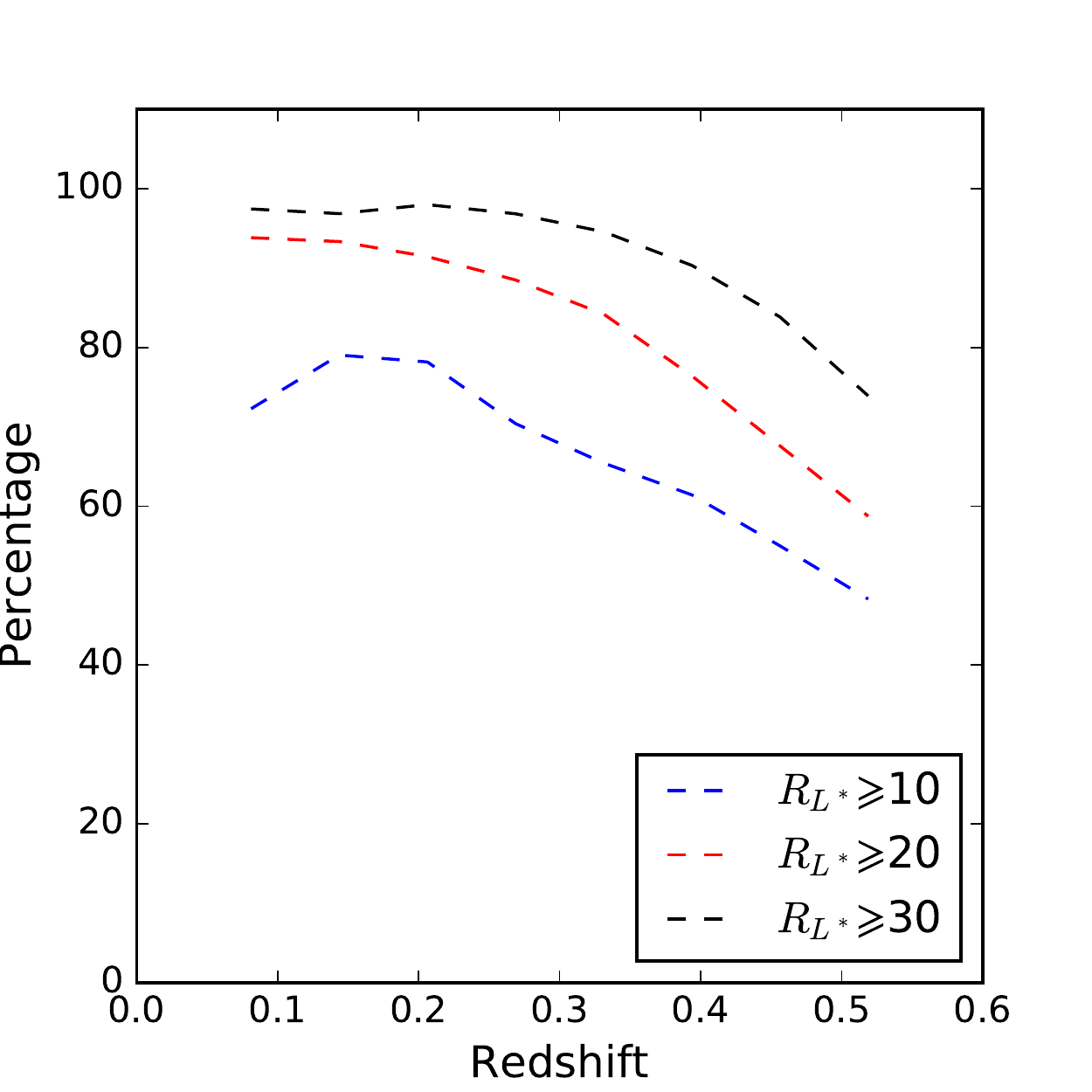}}		
		\caption{Left: Richness distributions of the matched and un-matched clusters in our catalog. Middle: richness comparison of the matched clusters between our ($R_{L*}$) and WHL2012 ($R_{L*,500}$) catalogs. Right: The percentage of our clusters that are matched with WHL2012 catalog as function of redshift. Different lines represent for clusters with different richnesses.}
		\label{fig-compwen}
		\end{figure*}

\begin{figure}[!ht]
		\plotone{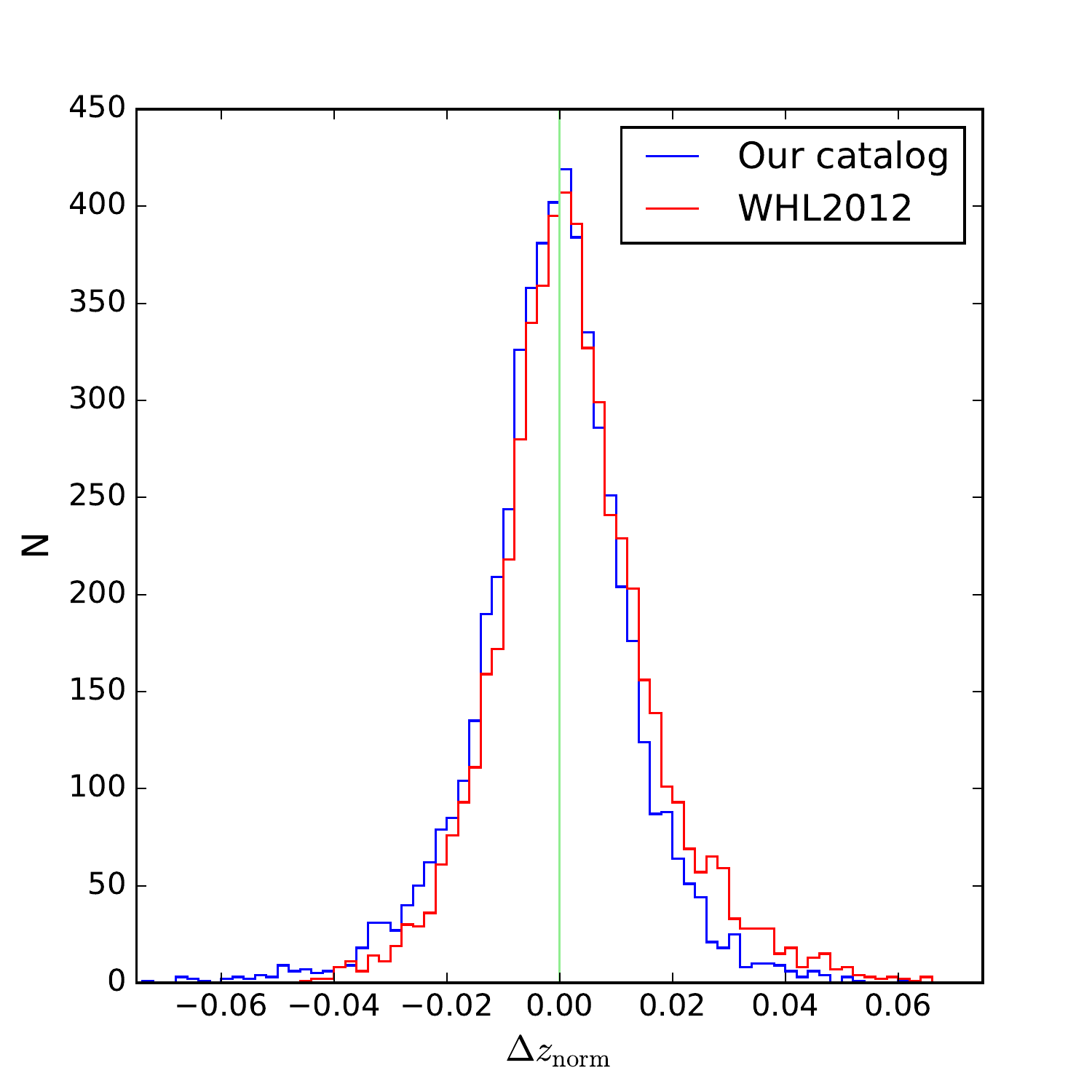}	
		\caption{$\Delta z_\mathrm{norm}$ distributions of matched clusters in our and WHL2012 catalogs. The vertical line shows $\Delta z_\mathrm{norm}=0$.}
		\label{fig-zcomp}
		\end{figure}

There are 32,884 galaxy clusters in the WHL2012 catalog that are located in our footprint and in the redshift range of $0.05 < z < 0.65$. These two catalogs are cross-matched using a redshift error of  $|\Delta z| \leqslant 0.05 (1+z)$ (approximately 2.5 times the photo-z error of clusters) and a separation of 2 Mpc. If spectroscopic redshifts are available, we use the spectroscopic redshifts instead of photo-zs for cross-matching. In addition, if the cluster in WHL2012 and corresponding cluster in our catalog have the same center (BCG), we consider them as an identical cluster regardless of the redshift difference. We take the WHL2012 catalog as the reference and get a total of12,049 matched clusters in our catalog (61.4\%). Since the cluster detection of WHL2012 is based on only SDSS photometric data and our work is based on both optical and near-infrared data, the number of galaxies used in WHL2012 is much more than that in this work. A considerable number of the WHL2012 clusters have relatively low richness so that they are below our detection thresholds. The left panel of Figure \ref{fig-compwen} shows the richness distributions of the matched and un-matched clusters in our catalog. The common clusters identified by both WHL2012 and this work have relatively large richnesses. Among the un-match 6,363 clusters, there are 1,598 clusters (25.1\%) that can be matched to the AMF catalog of \citet{ban2018}. Themiddle panel of Figure \ref{fig-compwen} shows the richness comparison of the matched clusters between our and WHL2012 catalogs. The richness estimations of  these two catalogs are consistent. The right panel of Figure \ref{fig-compwen} presents the fraction of our clusters that are also detected by WHL2012 as function of redshift. The fraction of matched clusters decreases as the redshift increases or richness decreases. For $R_{L*} \geqslant$ 10, 20, and 30, the matching rates are about 66.3\%, 78.2\%, and 89.0\%, respectively.  We also compare the redshift accuracy of our and WHL2012 cluster catalogs. A  sample of 5,488 matched clusters have spectroscopic redshifts. Figure \ref{fig-zcomp} shows the distributions of  $\Delta z_\mathrm{norm}$ for both catalogs. The standard deviation and bias of $\Delta z_\mathrm{norm}$ of these matched clusters in WHL2012 is 0.0127 and 0.0021, respectively, and those in our catalog are 0.012 and -0.0007, respectively. The cluster redshifts of our catalog has slightly smaller scatter and bias, but as seen in Figure 8c of \citet{gao2018}, our photo-z bias improves substantially relative to other deep photometric redshift catalog, when comparing the photo-z based on only SDSS photometry.  

\subsubsection{Comparison with the redMaPPer catalog}

\citet{ryk2014} introduced a red-sequence cluster finder, redMaPPer, and applied to the SDSS DR8 catalog. They presented a catalog of $\sim$25,000 clusters over the redshift range of $0.08 < z < 0.55$. \citet{roz2015} and \citet{ryk2016} updated the algorithm and provided updated catalogs. In this paper, we use the latest version (v6.3\footnote{\url{http://risa.stanford.edu/redmapper/}}) of the redMaPPer catalog for comparison, which consists of 26,311 clusters over the redshift range of $0.08 < z < 0.6$.

\begin{figure*}[!ht]
		\plottwo{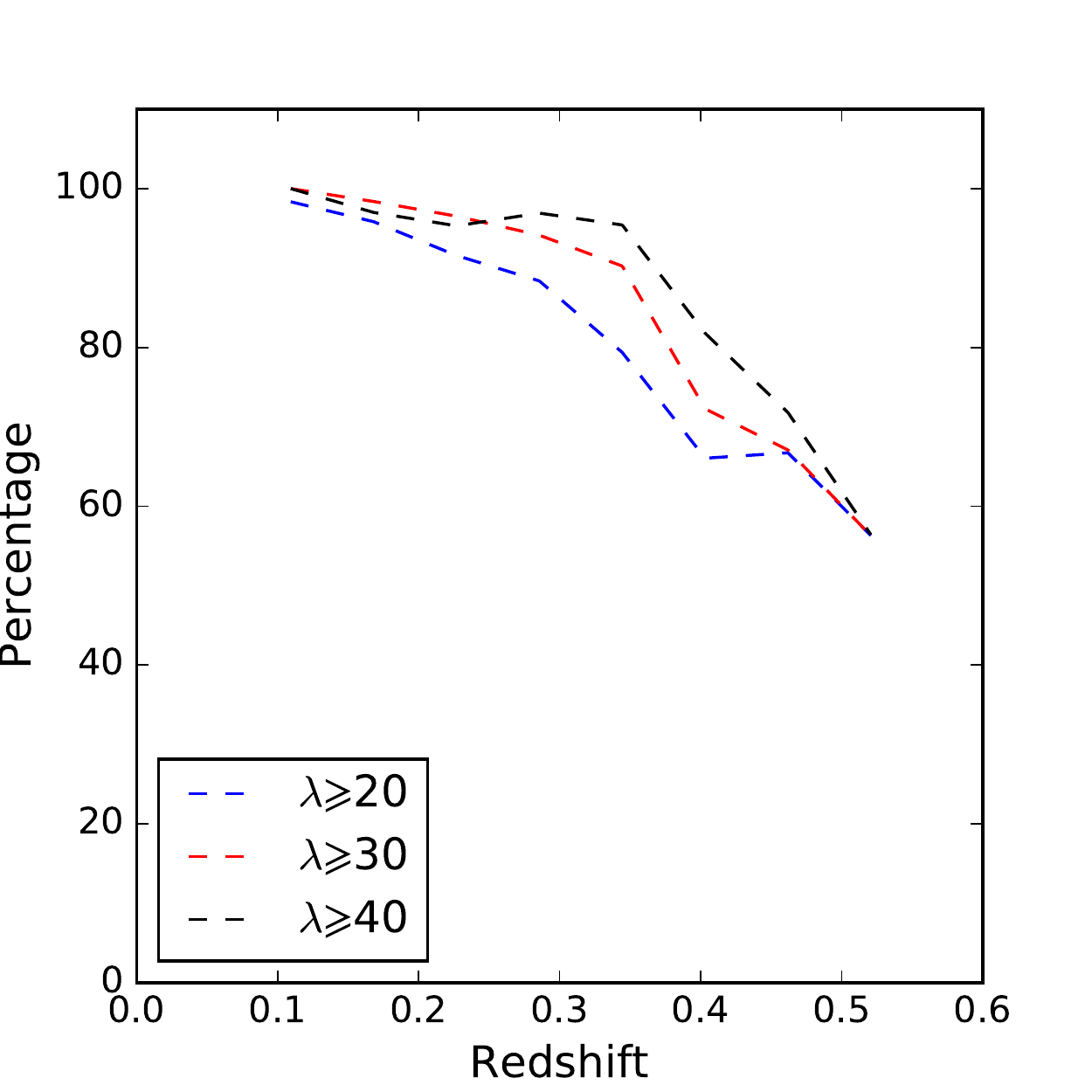}{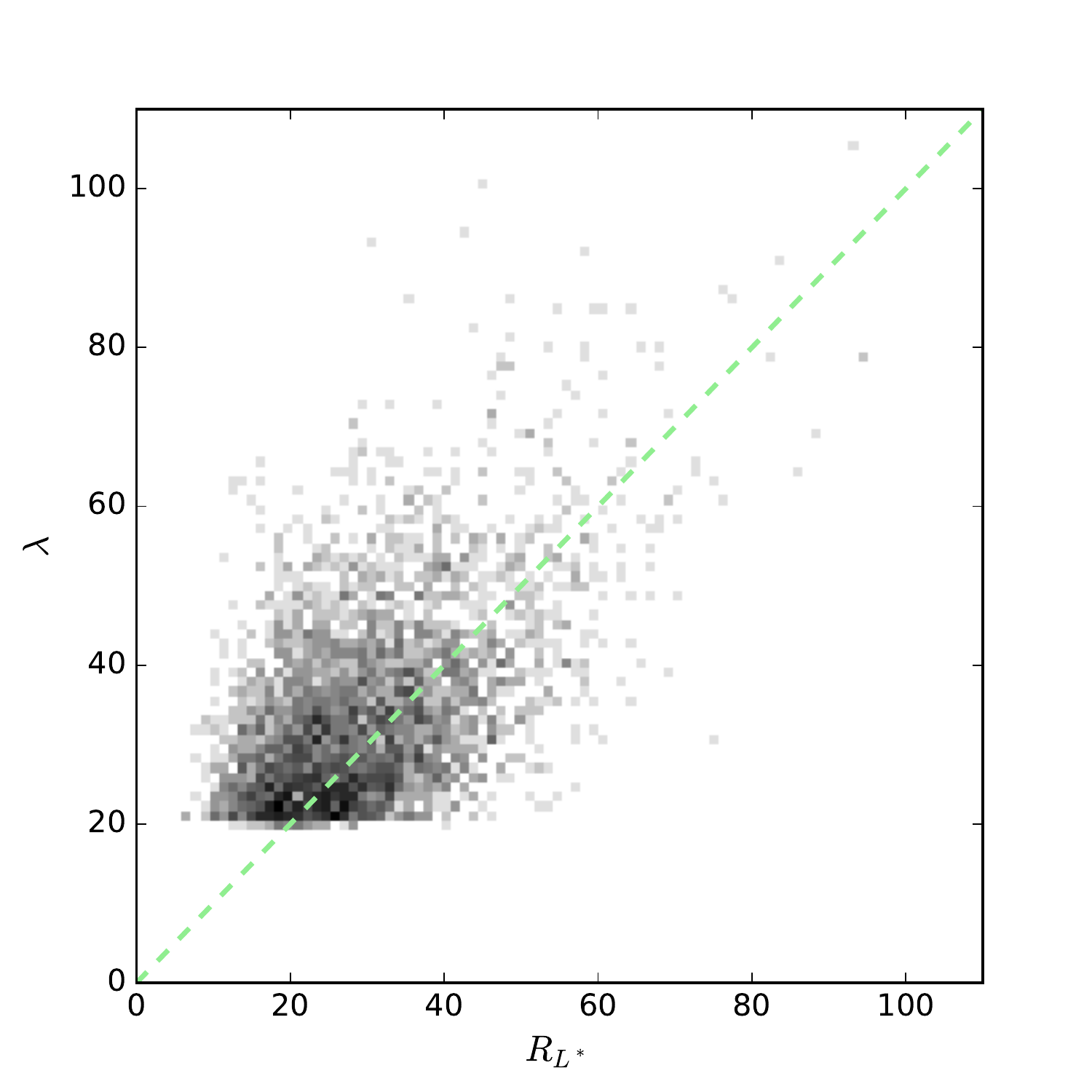}	
		\caption{Left: the percentage of redMapper clusters that are matched with our catalog for different richnesses as function of redshift. Different lines represent for clusters with different richnesses.  Right: richness comparison of the matched clusters between our ($R_{L*}$) and redMaPPer ($\lambda$) catalogs. The dashed line shows $y=x$.}
		\label{fig-compred}
		\end{figure*}

There are a total of 7,905 galaxy clusters in the redMapper catalog located in our footprint. A sample of 5,881 (74.4\%) redMapper clusters are matched with our catalog using a redshift error of  $|\Delta z| \leqslant 0.05 (1+z)$ and a separation of 2 Mpc. The left panel of Figure \ref{fig-compred} shows the fraction of the redMaPPer clusters that are also detected by us for different richnesses as function of redshift.  The matching rate increases with decreasing redshift and increasing richness. The redMaPPer cluster richness($\lambda$) is defined as the sum of the membership probabilities of all the galaxies within a scale-radius $R_{\lambda}$  \citep{ryk2012,ryk2016}. We compare the redMaPPer richness $\lambda$ with our $R_{L*}$ in Figure \ref{fig-compred}. From this figure, we can see that the two richness estimations are roughly consistent, although the relation is somewhat diffuse.

\section{Summary} \label{sec:sum}
Based on the imaging data from SCUSS, SDSS, and unWISE, we have obtained a photometric redshift catalog of about 23.1 million galaxies with $r < 22$ mag and $z<0.8$. Comparing those photo-z catalogs based only on SDSS data, our photo-z presents less bias and higher accuracy. In this paper, we make use of this photo-z catalog to detect clusters by introducing a new cluster-finding algorithm, CFSFDP, which can detect the overdensities of galaxies efficiently. This algorithm detect the clusters in the plane of local density of a galaxy vs. shortest distance from others with higher local densities. It can rapidly find the local density peaks and corresponding members. 

A total of19,610 clusters at $0.05 < z < 0.65$ are identified over the area of about 3,700 deg$^2$ in the south Galactic cap. The catalog can be accessed in the webpage\footnote{\url{http://batc.bao.ac.cn/~zouhu/doku.php?id=projects:scuss_clusters:start}}.  Monte-Carlo simulations present that the overall completeness is92\% and the completeness rises to95.8\% for $N_\mathrm{1Mpc} \geqslant 24$. The overall false detection rate is about8.9\% and the false rate drops to3.6\% for $N_\mathrm{1Mpc} \geqslant 24$. Comparing with the galaxies with spectroscopic redshifts, the redshift uncertainty of our clusters is estimated to be about 0.0125. The richness and mass of the clusters are calibrated using the methods of \citet{wen2015,wen2018}, which is based on the X-ray and SZ measurements. Our clusters have the median richness of 21.7 and median mass of $1.2\times10^{14} M_\sun$. About 98\% of the Abell clusters in the SCUSS footprint are matched to ours clusters. About 64\% of our clusters are matched to the WHL2012 clusters. The matching rate increases with increasing richness and decreasing redshift.  For $R_{L*} \geqslant 30$, the matching rate reaches up to about 89\%.We also make a comparison with the redMaPPer cluster catalog, about 74.4\% of the redMaPPer clusters are identified by this work. The richness estimations are roughly consistent, although these two catalogs are based on two distinct algorithms.

We thank Z.L. Wen, F. Yang and Y.H. Xu at National Astronomical Observatory of China for private discussions. This work is supported by the Chinese National Natural Science Foundation (grant Nos. 11433005, 11673027, 11733007), the National Basic Research Program of China (973 Program; grant Nos. 2015CB857004, 2014CB845704, 2014CB845702, and 2013CB834902), and the External Cooperation Program of Chinese Academy of Sciences (grant No. 114A11KYSB20160057).This work is also supported by Major Program of National Natural Science Foundation of China (No. 11890691).



\end{document}